\documentclass{aa}  

\usepackage{graphicx}
\usepackage{diagbox}
\usepackage{txfonts}
\usepackage{lipsum}
\usepackage{subcaption}         
                                
\usepackage{lscape}           
                                
\usepackage{placeins}          
                               
\usepackage[normalem]{ulem}
\usepackage{threeparttable}

\usepackage[colorlinks,allcolors=blue]{hyperref}
\defcitealias{baubock_rotational_2015}{B15}

\newcommand{\msun}{\mbox{$\,{\rm M}_\odot$}}
\newcommand{\rns}{\mbox{$R_{\rm NS}$}}
\newcommand{\mns}{\mbox{$M_{\rm NS}$}}
\newcommand{\mr}{\mns--\rns}
\newcommand{\rinf}{\mbox{$R_{\infty}$}}

\begin{document}

   \title{Mass and radius measurements of the neutron star 47~Tuc X7}
   \subtitle{A new bias-free method}

   \author{C. Kazantsev\inst{1}\fnmsep\thanks{Corresponding author: christine.kazantsev@utoulouse.fr}
        \and S. Guillot\inst{1}
        \and L. Mauviard\inst{1}
        \and T. Salmi \inst{2}
        \and N. A. Webb\inst{1}
        }

   \institute{Univ Toulouse, CNES, CNRS, IRAP, Toulouse, France
   \and Department of Physics, University of Helsinki, P.O. Box 64, FI-00014 University of Helsinki, Finland
   }
   
   \date{Received September 30, 20XX}

  \abstract
   {Neutron star (NS) radius measurements provide precious information to constrain the dense matter equation of state (EOS). Low-mass X-ray binaries in quiescence (qLMXBs) have been used for this purpose, but a number of sources of systematic biases were uncovered, making other sources (NS-NS mergers or millisecond pulsars, MSPs) more favored for EOS studies.}
   {We aim to reintroduce qLMXBs as reliable sources of NS mass and radius measurements, with a new method, free of systematic biases. We test our implementation on the qLMXB X7 in the globular cluster 47~Tucanae.} 
   {We used X-PSI, a software initially designed for pulse profile modeling of MSPs, to perform the spectral analysis of the 47~Tuc X7 observations. X-PSI accurately models the effects of the unknown NS rotation and possible surface anisotropies (two sources of biases in qLMXBs) on the NS spectra. The most significant source of bias on the radius is usually the chemical composition of the NS atmosphere, which, fortunately, in the case of 47~Tuc X7, is known to be hydrogen-rich.}
   {A broad range of masses and radii was explored, resulting in the characteristic curved M-R posterior contours of qLMXBs. We obtain a NS radius at 1.4\,\msun\ of $R_{1.4} = 12.9\pm0.4$\,km (68\% credible interval). A shift of the radius by less than a percent is measured compared to the model where these sources of systematic uncertainties are neglected. More importantly, including rotation and surface anisotropies in the modeling does not significantly broaden the radius posteriors. We also place strong constraints on the X-ray pulsed fraction (upper limit of 6.0\% at a 99.97\% credible level) caused by the possible presence of a hot spot. This suggests that, for 47~Tuc X7, robust radius constraints can be obtained even without considering systematics, likely because of the deep exposures. We use the resulting M-R constraints from this NS to quantify the improvement on an EOS inference when combined with other measurements.}
   {We show that, using recently developed tools, qLMXBs can be exploited to infer reliable NS masses and radii, which can in turn constrain the EOS.}

   \keywords{neutron stars --
                low-mass X-ray binaries}

   \maketitle
   \nolinenumbers

\section{Introduction}
\label{sec:intro}
Among the fundamental forces of nature, the strong force dominates the behavior of nuclear matter. Neutron stars (NSs) are unique probes to study the strong force, since their $\sim$1 -- 2\msun\ compressed into a $\sim$10\,km radius object leads to densities greater than that of atomic nuclei ($\rho_0 \sim 2.4\times10^{14}$~g/cm$^3$). A broad variety of equations of state (EOSs), i.e., the relation between pressure and density, have been theorized to quantify the repulsive forces of matter beyond saturation density $\rho_0$ \citep[for the current state of the art]{chatziioannou_2025}. The exact composition of NS cores remains unknown, ranging from pure nucleonic matter (protons, neutrons, and muons) to more exotic compositions such as pions or kaons condensates or even deconfined quarks (see review from \citealt{lattimer_neutron_2010} for details). Each EOS model corresponds to a unique relation between the mass, \mns, and the radius, \rns, of a NS. Precise measurements of \mns\ and \rns\ can thus yield crucial information on the EOS and the behavior of matter inside a NS.

Precise measurements of \mns\ can be obtained via pulsar timing of NSs in binary systems \citep{lorimer_2004}. Accurate determination of the orbital parameters (including those affected by general relativity) provides \mns\ with typical uncertainties of between a fraction of a percent for the most precise pulsar masses\footnote{\url{https://www3.mpifr-bonn.mpg.de/staff/pfreire/NS\_masses.html}} to a few tens of percent. The most massive NSs with a precisely determined gravitational mass have placed some constraints on dense matter by ruling out the EOSs that were predicting lower maximum \mns\ (e.g., \citealt{zdunik_2013}). However, a plethora of EOS models remain compatible with masses from 1.17\,\msun\ \citep{martinez_pulsar_2015} to 2.08\,\msun\ \citep{fonseca_refined_2021}.
Unless a mass higher than currently known is measured (which would exclude more EOS models), radius measurements provide a decisive tool to discriminate between theories of dense nuclear matter. 

Several methods exist to measure NS radii, either with rotation-powered millisecond pulsars (MSPs) or their previous evolutionary stage, low-mass X-ray binaries (LMXBs). Probing NSs at different stages of their evolution is important for an intercomparison of the results, either to check for possible biases inherent with each method, or to help understand possible effects of binary evolution on the NS internal structure.

A recent and now well-established method of measuring radii relies on pulse profile modeling (PPM) of the thermal emission from MSPs, old pulsars rotating at periods ranging from 1.4 to a few tens of milliseconds. These are typically observed with the Neutron Star Interior Composition Explorer (NICER; \citealt{gendreau_nicer_2016}) thanks to its time resolution. By modeling the general and special relativistic effects imprinted on the X-ray photons originating from surface hot spots on the rotating MSPs, this method has been able to provide measurements of \mns\ and \rns\ for five NSs: PSR~J0030+0451 \citep{vinciguerra_updated_2024, kini_nicer_2026}, PSR~J0437$-$4715 \citep{choudhury_nicer_2024, miller_radius_2025}, PSR~J0740+6620 \citep{salmi_radius_2024}, PSR~J1231$-$1411 \citep{salmi_nicer_2024}, and PSR~J0614$-$3329 \citep{mauviard_nicer_2025} being the most recent ones. When available, mass priors from radio timing are considered in the analyses. Masses derived from PPM without radio timing priors are typically less precise (about $\pm10$\% at best, for PSR~J0030+0451). Radius measurements have uncertainties in the range of $\pm$7\% to $\pm$15\% for all sources.  

Measurements of \mns\ and \rns\ have also been performed using LMXBs (e.g., \citealt{heinke_hydrogen_2006, guillot_measurement_2013}). These are binary systems with a compact object (here a NS) and a low-mass companion (typically a subsolar main-sequence star or a white dwarf). Over the past 20 years, the NS surface emission in LMXBs has been used extensively for such constraints (see below), typically when the accretion was reduced or suppressed. However, several sources of uncertainties associated with this method of measuring \mns\ and \rns\ (see Sect.~\ref{sec:biases}) were highlighted in the literature such that LXMBs were put aside in favor of PPM of MSPs with NICER, which was thought to provide more reliable constraints. 

Known LMXBs alternate phases of weeks to month-long accretion outbursts (with $L_X \sim 10^{36} - 10^{38}$\,erg/s; see \citealt{heinke_outburst_2025}), thought to be due to thermal or viscous instabilities in the disk (e.g., \citealt{lasota_2001}), and extended months to years of quiescence ($L_X \sim 10^{31} - 10^{33}$\,erg/s). In quiescence, if accretion is low enough or fully suppressed, the surface becomes visible, provided that the NS has been heated sufficiently by the accretion of material \citep{brown_crustal_1998}. Among those, quiescent LMXBs (qLMXBs) in globular clusters (GCs) have been used to measure \mns\ and \rns. In addition to having a precise distance, they present the advantage of remarkable flux stability ($<1-2$\% variability) over up to several decades (\citealt{bahramian_limits_2015} and references therein). This suggests the absence of ongoing low-level accretion in these systems. Their spectra can be modeled as a simple thermal spectrum with no power-law component (that could otherwise be explained by residual accretion). The most recently published works present the measurements for sources X5 and X7 in 47~Tucanae (47~Tuc, \citealt{bogdanov_neutron_2016}), in M13 \citep{shaw_radius_2018}, and in M30 \citep{echiburu_spectral_2020}. All are consistent with radii in the range 10 -- 15\,km, but specific values vary with each source and model. 

These results were obtained by spectroscopy of their high signal-to-noise ratio (S/N) thermal emission to measure the observed temperature, $T_{\infty}$, and size \rinf\ of the emitting region, which relate to the true size, \rns, and mass, \mns, and effective temperature, $T_{\rm eff}$, of the NS following the equations: 
\begin{align}
    \rinf &= \rns (1+z)= \rns \left( 1 - \frac{2G\mns}{\rns c^2}\right)^{-1/2}, \\
    T_{\infty} &= T_{\rm eff} / (1+z),    
\end{align}
with $(1+z)$ being the gravitational redshift. Measuring \rinf\ and $T_{\infty}$ can thus provide constraints on \mns\ and \rns. However, with three parameters and two observables, degeneracy between the parameters remains, resulting in strong correlations. In the \mr\ plane, the constraints roughly follow curves of constant \rinf, leading to an elongated, curved shape. Because of these correlations, statistical analyses combining measurements from a select number of qLMXBs have been done to extract constraints on the EOS \citep{baillot_detivaux_new_2019, lattimer_neutron_2014}.

However, for qLMXBs, these measurements rely exclusively on modeling their spectra, and any effects altering the spectral shape will bias the resulting parameters. A number of sources of biases (resulting in possible systematic uncertainties) were uncovered that affect the measurements of $T_{\infty}$ and \rinf\ and, by extension, \mns\ and \rns\ (see details in Sect.~\ref{sec:biases}). 
In this work, we present a new method of measuring \mns\ and \rns\ in qLMXBs, aiming to take into account the main sources of uncertainties to provide more reliable constraints. In Sect.~\ref{sec:biases} we describe the existing biases and their expected effects on the radius. In Sect.~\ref{sec:47Tuc} we present the source that will be the subject of the analysis, X7 in the GC 47~Tuc. In Sect.~\ref{sec:method_xpsi} the method of incorporating the biases in the \mns\ -- \rns\ inference is described, and the results are shown in Sect.~\ref{sec:results}. Finally, we discuss the results and perform an EOS inference in Sect.~\ref{sec:discussion} and present the conclusions in Sect.~\ref{sec:ccl}.

\section{Biases and sources of uncertainties}
\label{sec:biases}
In this section we summarize the known biases linked with the analyses of qLMXBs that have been reported in the literature and how these affect \rns\ measurements. 

\subsection{Distance to the source}
The most direct source of uncertainty is related to the distance to the qLMXB. Since the measured X-ray flux of a thermal blackbody-like component scales with $(\rinf/d)^2$, precise distances are required for accurate radius measurements. Any uncertainty on $d$ will directly reflect as an uncertainty on $\rinf$.  Historically, spectral analyses of qLMXBs in GCs have provided good measurements of $\rinf$ as their distances are well known with uncertainties of $\sim$ 5 -- 10\%. More recently, the latest \textit{Gaia} data release, which includes observations of GCs, allowed distances with accuracies better than 1\% to be derived. \citet{baumgardt_accurate_2021} in particular performed a systematic study of several GCs to derive highly accurate distance measurements. 

\subsection{Rotation of the neutron star}
\label{sec:spin}
None of the GC qLMXBs have a measured spin frequency, mainly because the time resolution of the observations performed is insufficient. A timing mode \textit{Chandra} High Resolution Camera (HRC, with 16\,$\mu$s resolution) exposure of 47~Tuc X7 only resulted in an upper limit on the pulsed fraction \citep{elshamouty_impact_2016}. Some accreting LMXBs do have spin measurements, and by analogy, GC qLMXBs are expected to have frequencies in the same range.

The spin frequency of NSs in accreting LMXBs (during an outburst) has been measured in some cases; for example, in Type I thermonuclear X-ray bursts. These bright X-ray bursts are caused by the unstable burning of accreted matter on the NS surface. They sometimes present short-lived oscillations during the burst rise or decay, which may be interpreted as the spreading burning front. Frequencies in the range of 200 -- 600\,Hz have been measured and are thought to be associated with the NS spin frequency (\citealt{galloway_thermonuclear_2021} or \citealt{stum_detection_2026} for a recent example). 
In accreting millisecond X-ray pulsars (AMSPs), the accretion material is channeled by the NS magnetic field onto the poles, leading to X-ray pulsations as the NS rotates. Spin frequencies measured from these pulsations range from 180 to 600\,Hz \citep{patruno_accreting_2021}.

The usual assumption for qLMXB spectroscopic analyses has been that of a nonrotating, spherically symmetric NS, because 1) no NS spin frequency was known for GC qLMXBs, and 2) there is a lack of models and tools to account for rotational effects in the spectral fit. Nonetheless, it was shown in \citet[hereafter \citetalias{baubock_rotational_2015}]{baubock_rotational_2015} that rotation does have an impact on the inferred radius. The spectrum of a rapidly spinning NS is broadened by the Doppler effect compared to that of a non-spinning one. Photons emitted from the side moving toward the observer will be blueshifted, while those from the side moving away will be redshifted. Additionally, because of rotation, the NS shape will become slightly oblate, which is an EOS-dependent effect. Both the Doppler effect and oblateness will broaden and shift the spectrum, and neglecting such effects underestimates the radius. The magnitude of the bias depends on a number of parameters, the main ones being radius, spin frequency, and inclination angle, and is stronger for high spin frequencies. As an example, for a NS spinning at 600\,Hz, \citetalias{baubock_rotational_2015} estimated the bias on the spectroscopic radius \rinf\ to be a few percent on average, going up to $12$\% for a 15\,km-NS observed edge-on.

\subsection{Surface emission anisotropy}
Any emission anisotropy from the surface would manifest itself as a variation in the X-ray flux with rotation, i.e., pulsations, as seen in thermally emitting MSPs. So far, no X-ray pulsations have been discovered in GC qLMXBs, either because 1) the observations did not have the time resolution to be sensitive to the expected NS rotation spin (see Sect.~\ref{sec:spin}), 2) there is insufficient contrast between areas at different temperatures, due to a low spectral sensitivity, a small temperature differential, or low photon collection, 3) the anisotropies are axially symmetric (e.g., an equatorial ring), or 4) the surface is emitting isotropically. The latter is the most likely explanation in the context of deep crustal heating \citep{brown_crustal_1998}. Indeed, the source of heat in qLMXBs finds its origin deep in the crust that has been heated by reactions resulting from past episodes of accretion \citep{haensel_models_2008}, such that, during quiescence, the heated crust re-radiates the accumulated heat uniformly in all directions through the outer NS layers.

This led to the usual assumption of a uniform surface emission in qLMXBs. But since no firm demonstration of this assumption has even been made to our knowledge, either observationally or theoretically, surface inhomogeneities could be present. In this case, a second thermal component at a higher temperature (e.g., a hot spot) would be present in the observed spectrum. However, qLMXB spectra are generally well fit by a single-temperature thermal component, without the need for an additional one. Neglecting this potential second component would bias the temperature toward higher values, and consequently the radius would be biased to lower values. 

\citet{elshamouty_impact_2016} studied the effect of adding a second thermal component (for a hot spot) in the modeling. Combining \textit{Chandra} Advanced CCD Imaging Spectrometer (ACIS)-S observations for a phase-averaged spectrum with \textit{Chandra} HRC-S data for a phase-resolved light curve, they investigated the possibility of a hot spot being present and not detected. By varying various relevant parameters (frequency, compactness, and hot spot parameters), they showed that the pulsed fraction can be constrained to be lower than $\sim 10-30$\% (depending on the sources) and quantified the possible downward bias on the radius to be as large as $\sim30$\%.

\subsection{Atmosphere composition}
The atmosphere of the NS is the outermost layer of its surface (last $\sim$\,centimeter, \citealt{potekhin_atmospheres_2014}) and the layer in which the last scattering of photons takes place, resulting in the observed emergent spectrum. For NS in accreting systems, this atmosphere is composed of the matter previously accreted from the companion star. Because of the intense surface gravity, stratification of the accreted material happens on 10 -- 100\,s timescales \citep{bildsten_fate_1992}, leaving the lightest element, usually hydrogen or helium, in the outermost layer. Given the temperatures of $T\sim10^6$\,K and low magnetic fields (see Sect.~\ref{sec:bfield}), this atmospheric layer is expected to be effectively fully ionized \citep{zavlin_model_1996}. 

Since the NS atmosphere composition depends on the accreted material, the nature of the companion donor star is key to determining the NS atmospheric composition. A main-sequence companion will lead to a hydrogen NS atmosphere, while a helium (or heavier elements) atmosphere could be considered for white dwarf companions or other H-depleted stars. The radiative transfer in the atmosphere depends on the composition, via the opacities of the elements present, such that different models must be used for spectral analyses. For qLMXBs, this is the largest source of biases in radius measurements, as the choice of one atmosphere model over another can change the inferred radius by up to $\sim 50$\%. For example, it was demonstrated that He-atmosphere models predict higher radii than H-atmosphere models for some qLMXBs (see, for example, \citealt{servillat_M28_2012, catuneanu_M13_2013, heinke_improved_2014, bogdanov_neutron_2016, shaw_radius_2018}). 

\subsection{Magnetic field}
\label{sec:bfield}
Estimating the magnetic fields of NSs generally requires one to either know the spin and spin down and assume a dipolar field \citep{lorimer_2004}, or to measure the energy of a cyclotron resonant scattering feature in the hard X-ray spectrum \citep{staubert_cyclotron_2019}. Since none of those are available for GC qLMXBs, their magnetic fields are unknown. Yet, by analogy with LMXBs (in the field of the Galaxy), one can infer magnetic fields on the order of $\sim 10^8 - 10^9$\,G for qLMXBs. Indeed, AMSPs have been used to measure the strength of the magnetic field of LMXBs; for example, by \citet{mukherjee_magnetic-field_2015}, which used information on the flux variations during pulsations to constrain the magnetic dipolar moment. Focusing on 14 AMSPs, they found that $B \sim 10^8$\,G, with the B field estimated over all sources to have an upper limit of $B \leq 4\times 10^9$\,G. These low magnetic field values, in a similar range to MSPs, can be explained by the advection of the magnetic field lines below the surface by the accreted material, as was explained in \citet{romani_unified_1990}. 

At such low magnetic field strengths, the radiative transfer in the NS atmosphere remains unaffected. Indeed, for surface temperatures of $T\sim10^6$\,K (measured in qLMXBs), the opacity of free-free interactions remains the same for magnetic fields lower than $B\sim 10^{10}$\,G \citep{zavlin_model_1996}, as the photon frequency is larger than the electron cyclotron frequency. Since this condition is verified in qLMXBs, radiative transfer models can thus assume nonmagnetic atmospheres, and magnetic fields can safely be neglected without affecting the results.

\subsection{Photon pileup}
\label{sec:pileup_bias}
X-ray imagery, for example with the \textit{Chandra} ACIS camera, has been known to suffer from photon pileup when observing bright sources. This instrumental effect happens when two (or more) photons arrive on the same pixel of the CCD camera at the same time (i.e., within a single readout time, up to 3.2\,s with the \textit{Chandra} ACIS detector). The resulting detected event will either be rejected as a ``bad" event, i.e., filtered out with proper flagging, or kept as a ``good" event, but with the energy of the sum of the individual photon energies. \citet{bogdanov_neutron_2016} showed that even a 1 -- 2\% pileup fraction can shift by $\sim 8-10$\% the inferred radius if it is not included in the modeling (for a 1.4\,\msun\ NS; see their Fig.~3). Higher fractions of pileup will lead to even higher shifts ($\geq$ 50\% shift for an uncorrected 10\% pileup fraction, based on \texttt{XSPEC}). Pileup will harden the observed spectrum, yielding a higher temperature and a smaller radius than the true values.
Small amounts of pileup can be corrected in the spectral analyses with \texttt{XSPEC} \citep{arnaud_xspec_1996} or \texttt{ISIS} \citep{houck_isis_2000} thanks to the implemented model of \citet{davis_event_2001}.

\section{47~Tuc X7}
\label{sec:47Tuc}
\subsection{Description}
\label{sec:47Tuc_desc}

The source of interest in this work is the qLMXB X7, one of the confirmed qLMXBs in the GC 47~Tuc (also known as NGC~104), located $d=4.52\pm0.03$\,kpc from Earth \citep{baumgardt_accurate_2021}. X7 has the brightest X-ray flux of all known GC qLMXBs and an extensive \textit{Chandra} dataset containing 21 ACIS observations for a total 560\,ks exposure. As for all GC qLMXBs, it displays no flux variations on timescales of years \citep{bogdanov_neutron_2016} and no evidence of residual accretion. Its spectrum appears to be purely thermal (with seemingly one single thermal component). Previous work incorporating a nonthermal component found power-law normalizations consistent with zero \citep{bogdanov_neutron_2016}. The first observations date from 2000--2002 and are known to suffer heavily from pileup, since they were obtained in full-frame mode (readout time, or frame time, of 3.2\,s). Subsequent observations from 2014 were performed in the 1/8 sub-array mode, reducing the frame time to 0.4\,s. This significantly reduced the probability of pileup events, from 10\% to 1\%\footnote{Estimated with WebPIMMS: \url{https://cxc.harvard.edu/toolkit/pimms.jsp}}. 

As most bright qLMXBs have, this source has already been studied in the context of \mr\ measurements and EOS inference \citep{heinke_hydrogen_2006, bogdanov_neutron_2016, baillot_detivaux_new_2019}. The work from \citet{bogdanov_neutron_2016}, which used exclusively the low pileup \textit{Chandra} dataset from 2014 -- 2015, obtained a radius of 11.0$\pm$0.8\,km for a 1.4\msun\ NS assuming a H-atmosphere.

Recently, a study of archival \textit{HST} observations of 47~Tuc was performed to identify the optical counterparts of several X-ray sources, including X7 \citep{berg_simultaneous_2024}. For the latter, the likely companion (named N1) is a main-sequence star with a weak H$\alpha$ excess originating from an accretion disk. N1 being hydrogen-rich, a NS with a hydrogen atmospheric composition is the preferred hypothesis and is used in this work. As the second most likely companion (N4) is also hydrogen-rich, the choice of atmosphere model appears to be reliable.

\subsection{Data reduction}
\label{sec:data_Chandra}
In the analysis presented here, we used all archival \textit{Chandra} data of this cluster from 2000 to 2022, obtained with the ACIS-I and ACIS-S cameras. The data were reprocessed using the \texttt{chandra\_repro} task of CIAO v4.17 \citep{fruscione_ciao_2006} with the CALDB v4.12, and the source and background spectra were extracted using \texttt{specextract}\footnote{Despite analyzing the spectra over the 0.3 -- 4.0\,keV energy range, the extraction was performed using the\texttt{specextract} option \texttt{energy=0.25:11.0:0.01} such that the response matrices are created over that energy range and photons with an incident energy below 0.3\,keV that would be redistributed above 0.3\,keV are accounted for.}. The source spectrum was extracted from a 2.5\arcsec circle centered on X7 (RA=$00^h24^m03^s528$, DEC=$-72\degr04\arcmin51\arcsec6$), corresponding to $\sim$95\% of the enclosed energy fraction. The background was extracted from an annulus ($r_\mathrm{in}=$3\arcsec, $r_\mathrm{out}=$8\arcsec) centered on a nearby source (X5) to get a source-free region. To reduce the number of observations to load for the analyses (see Sect.~\ref{sec:method_xpsi}) and the computation time, observations obtained with the same instrument and settings were combined using \texttt{combine\_spectra}, giving a total of seven datasets. All observations and relevant parameters are reported in Table~\ref{tab:data}. 

\begin{table}[t]
    \centering
    \begin{threeparttable}
    \caption{\textit{Chandra} observations of the GC 47~Tuc used in this work.}
    \begin{tabular}{c|c|c|c|c}
         Instrument & ObsID & Date & Exposure (s) & $\tau$ (s) \\
         \hline \hline
         ACIS-I & 78 & 03/2000 & 3875 & 0.9 \\
         ACIS-I & 956 & 03/2000 & 4691 & 0.9 \\
         \hline
         ACIS-I & 953 & 03/2000 & 31676 & 3.2 \\
         ACIS-I & 955 & 03/2000 & 31676 & 3.2 \\
         \hline
         ACIS-I & 954 & 03/2000 & 845 & 0.5 \\
         \hline
         ACIS-S & 2735 & 09/2002 & 65238 & 3.1\\
         ACIS-S & 2736 & 09/2002 & 65244 & 3.1\\
         ACIS-S & 2727 & 10/2002 & 65241 & 3.1\\
         ACIS-S & 2738 & 10/2002 & 68772 & 3.1\\
         \hline
         ACIS-S & 3384 & 09/2002 & 5307 & 0.8\\
         ACIS-S & 3385 & 10/2002 & 5307 & 0.8\\
         ACIS-S & 3386 & 10/2002 & 5545 & 0.8\\
         ACIS-S & 3387 & 10/2002 & 5735 & 0.8\\
         \hline
         ACIS-S & 15747 & 09/2014 & 50035 & 0.4\\
         ACIS-S & 15748 & 10/2014 & 16238 & 0.4\\
         ACIS-S & 16527 & 09/2014 & 40875 & 0.4\\
         ACIS-S & 16528 & 02/2015 & 40284 & 0.4\\
         ACIS-S & 16529 & 09/2014 & 24700 & 0.4\\
         ACIS-S & 17420 & 09/2014 & 9132 & 0.4\\
         \hline
         ACIS-S & 26229 & 01/2022 & 9650 & 3.0\\
         ACIS-S & 26286 & 01/2022 & 9827 & 3.0\\
         \hline
    \end{tabular}
    
    \label{tab:data}
    \tablefoot{Spectra within each horizontal lines have the same frame time, $\tau$, and were combined for the analysis.}
  \end{threeparttable}
\end{table}

\begin{figure}
    \centering
    \includegraphics[width=1\linewidth]{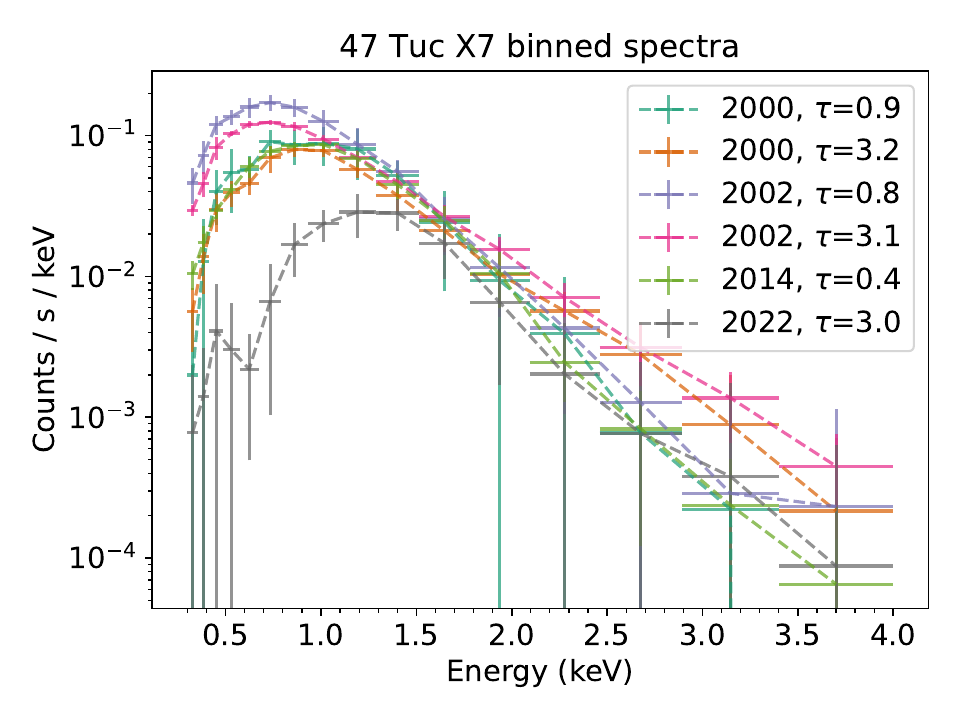}
    \caption{Spectra of the combined observations listed in Table~\ref{tab:data}. For visualization purposes, the spectra have been grouped in 16 bins. Year and associated frame time, $\tau$, are indicated in legend. ObsID 954 is not represented here as it is a short observation with a low number of counts.}
    \label{fig:spectra}
\end{figure}

The spectra of the combined data files are shown in Fig.~\ref{fig:spectra}, where they are grouped into 16 bins over the 0.3 -- 4.0\,keV range for visualization purposes only. The spectral analysis described in Sect.~\ref{sec:results} was performed on unbinned data. 
While Fig.~\ref{fig:spectra} could let one think that the source flux varied over two decades, the observed differences stem from the use of different detectors, their respective sensitivities, and pileup effects.

\section{Controlling the sources of uncertainties -- The X-PSI implementation}
\label{sec:method_xpsi}
As described in Sect.~\ref{sec:biases}, two major sources of uncertainties in qLMXB spectroscopy are related to the possible anisotropy of the emission and to the rotation of the NS. 
Accounting for NS rotation and a nonuniform surface in the modeling of NS emission requires ray-tracing capabilities in a general relativistic metric (e.g., see \citealt{elshamouty_impact_2016, baubock_rotational_2015} in the context of qLMXBs). In the present work, we employed the software X-PSI\footnote{\url{https://github.com/xpsi-group/xpsi}} (X-ray Pulse Profile Simulation and Inference, \citealt{riley_x-psi_2023}), which was developed and used primarily in the context of the pulse profile analyses of MSPs observed with NICER, i.e., for phase-resolved spectroscopy to infer \rns. X-PSI models gravitational and rotational effects on the NS surface emission, using the oblate Schwarzschild approximation to describe the spacetime surrounding the NS \citep{algendy_universality_2014}. Originally developed for the hot spot emission of various shapes, it also allows one to model the emission from the rest of the NS surface. In this section we present our adaptation of X-PSI in the context of qLMXB spectral analyses to include the effects of the NS rotation on the spectrum and the possibility of a hot region. Full details on X-PSI procedures and methods will not be described here but can be found in \citet{riley_nicer_2019}\footnote{All publications using X-PSI are listed here \url{https://xpsi-group.github.io/xpsi/applications.html}.}.

\subsection{X-PSI implementation}
\subsubsection{Phase resolved versus phase invariant modeling and the \texttt{everywhere} module}
In MSPs, one or several hot spots are observed, leading to a periodic light curve as the hot spots come in and out of view. Thanks to NICER's time resolution ($\sim$100\,ns), the X-ray data can be folded at the (known) period of the pulsar to produce phase-resolved spectra, which can then be analyzed with X-PSI to infer relevant parameters, including \mns\ and \rns\ \citep{bogdanov_constraining_2019_paperI}.

However, the \textit{Chandra} ACIS data of 47~Tuc used in our work (Sect.~\ref{sec:data_Chandra}) do not a have sufficient time resolution ($\tau$=0.4\,s at best) to detect possible pulsations at the expected spin frequencies of LMXBs, such that only phase-averaged data can be produced\footnote{Time-resolved (16\,$\mu$s resolution) data of 47~Tuc have been obtained with HRC in 2005 -- 2006, but they do not have the spectral resolution needed for the spectral fitting method described here ($\Delta E/E \sim 1$). Consequently, this HRC dataset is not considered in this work, and only the ACIS dataset will be used.}. Therefore, potential hot spots or nonuniform emission can only be detected spectrally. In all previous studies of qLMXBs, a single thermal component emanating from the whole surface of the NS was used to fit the data. 

In the present paper, we explore a few different emission scenarios. The ``default" one is that of a uniform surface emission, i.e., a single thermal component. This emission is implemented in X-PSI in a Python module called \texttt{Everywhere} (as opposed to \texttt{HotRegion} for modeling hot spots). This module works similarly to \texttt{HotRegion}, but for a signal with only one phase bin. To take advantage of this and speed up the computation, a different integrator for the surface of the NS is used, \texttt{integrator\_for\_time\_invariance}. Using the regular integrator (of \texttt{HotRegion}) would be computationally inefficient for a time-independent signal.  

This \texttt{Everywhere} module of X-PSI has not been tested in the literature, given the use of this software for MSPs with hot spot emissions. Therefore, we compared it to a setup designed to mimic the \texttt{Everywhere} but using modules that have been robustly tested: \texttt{HotRegion} and \texttt{Elsewhere} (e.g., \citealt{vinciguerra_updated_2024}). To do so, we generated a uniform NS surface using a single-temperature \texttt{HotRegion} (a hot spot at temperature $T_\mathrm{spot}$) combined with an \texttt{Elsewhere} component (to model the emission of the rest of the surface) and enforcing $T_\mathrm{else}=T_\mathrm{spot}$. The computed likelihoods in each configuration were identical, which demonstrated that the \texttt{Everywhere} module behaves as expected. Using the \texttt{integrator\_for\_time\_invariance} provided a speedup in likelihood computation by a factor of 14. In the process, a select number of minor issues were corrected in X-PSI v3.0.5 (details can be found in the corresponding changelog\footnote{\url{https://github.com/xpsi-group/xpsi/tags}}).

\subsubsection{The pileup module}
\label{sec:pileup}
\begin{figure*}[t]
    \centering
    \includegraphics[width=1\linewidth]{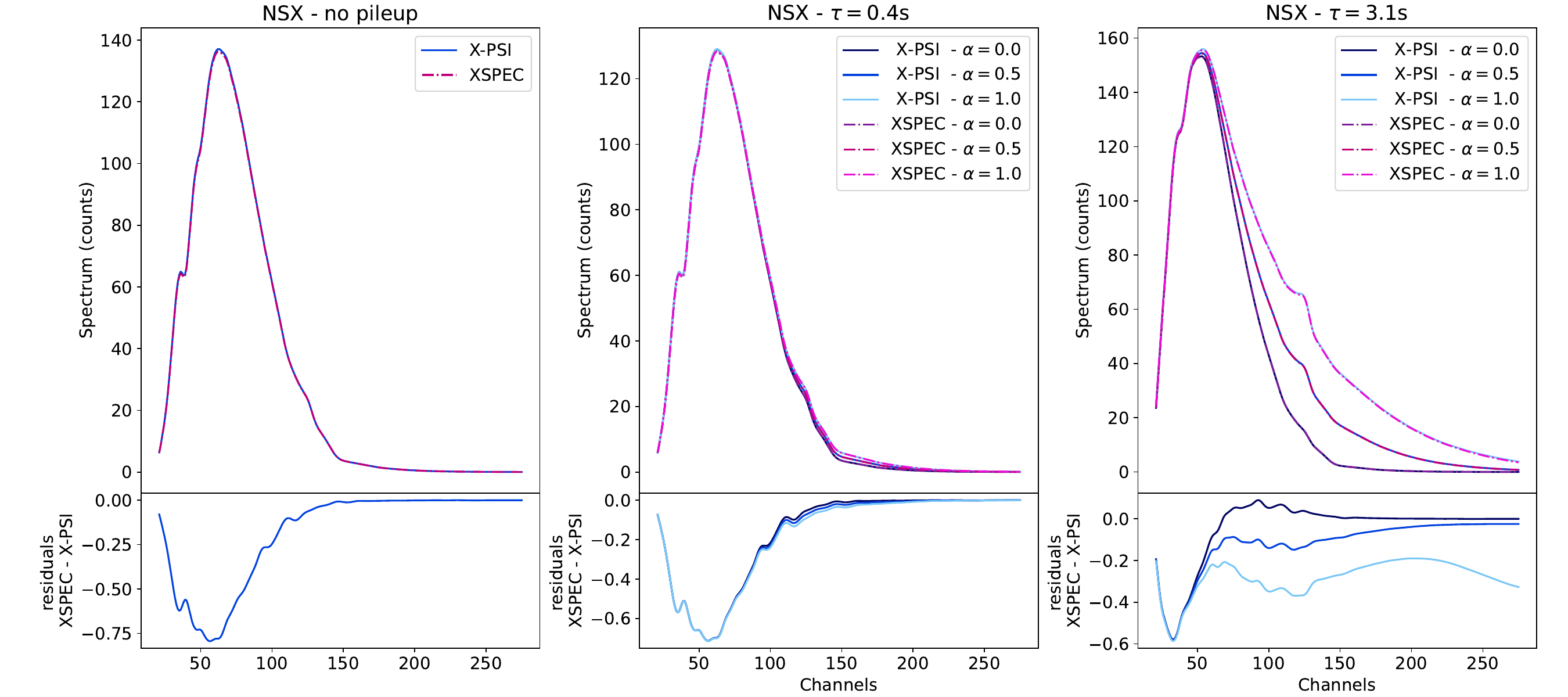}
    \caption{\textit{Left panel:} \texttt{XSPEC} and X-PSI predicted models in counts per channel with a \texttt{nsx}-Hp atmosphere model and no pileup. \textit{Center and right panels:} Same but with a pileup computation. The center panel is from observation 15747 with a frame time of 0.4\,s, while the right panel is from observation 2738 with a frame time of 3.1\,s. In all panels, X-PSI is in solid lines and \texttt{XSPEC} is shown with dashed lines. The residuals (\texttt{XSPEC} $-$ X-PSI) in counts are in the bottom panels.}
    \label{fig:compa_pileup}
\end{figure*}
Pileup was incorporated into a dedicated model in X-PSI, using the implementation of \citet[][in particular Eq.~A4 therein]{davis_event_2001}. It assumes that photon arrival times are Poisson-distributed and describes the total number of counts affected by pileup as a convolution product summed over the number of piled-up events (number of photons arriving within a frame time). In a scenario not affected by pileup, counts per channel are computed by multiplying the energy spectrum S(E) by the ancillary response files (ARF) and the redistribution matrix files (RMF): $\rm{RMF} \times \rm{ARF} \times S(E)$. The use of the pileup model imposes that one must perform the aforementioned convolution product after ARF multiplication but before the RMF: $\rm{RMF} \times Pileup \left[\rm{ARF} \times S(E)\right]$; see \citealt{davis_event_2001} for the full equation. 
In X-PSI v3.1, the \texttt{Instrument} class was modified to incorporate these changes into a new \texttt{InstrumentPileup} subclass, which makes use of the new module \texttt{PileupModule} where all the \citet{davis_event_2001} computations are implemented. 

To validate the pileup model in X-PSI, we first compared thermal emission spectral models (without data) between X-PSI and commonly used spectral analysis softwares with pileup implementation (XSPEC v12.14.1, which is used in the following, and ISIS\footnote{Results with ISIS were identical to those of XSPEC.}). We selected the instrument response files from two ACIS observations from 47~Tuc (obsIDs 2738 and 15747), which have two very different frame times (3.1\,s and 0.4\,s, respectively). 

We present our validation analysis with NS atmosphere models. The only NS H-atmosphere model common to both X-PSI and \texttt{XSPEC} is the \texttt{nsx} atmosphere model with partially ionized hydrogen (\texttt{nsx}-Hp in the following, \citealt{ho_neutron_2009}). In \texttt{XSPEC}, it corresponds to setting \texttt{specfile=1}, while in X-PSI it is in the form of precomputed intensity tables (see \citealt{salmi_atmospheric_2023}). 
We generated spectra (in counts per channel) in \texttt{XSPEC} and X-PSI with the same parameters for the \texttt{nsx}-Hp model, initially without pileup (left panel of Fig.~\ref{fig:compa_pileup}). 
Small differences $<1\%$ persist (see the residuals panel), which likely come from the implementation of the instrument responses in \texttt{XSPEC} and X-PSI. Computing the emergent photon flux spectra (i.e., before the instrument response is accounted for) yielded differences at the $<0.1\%$ level between X-PSI and \texttt{XSPEC}, seemingly numerical noise.

The pileup modeling could then be added, and the differences between the \texttt{XSPEC} and X-PSI implementations can be seen on the center and right panels of Fig.~\ref{fig:compa_pileup}. Varying either frame time (by choosing one observation or the other, center or right panel) or grade migration (by varying the corresponding parameter, different colored lines) still gives an excellent agreement, with less than 0.5\% difference in the predicted counts. 
Using a blackbody atmosphere model gives similar results to the ones of Fig.~\ref{fig:compa_pileup}.

\subsection{Comparison analysis with XSPEC}  
\label{sec:bogd17}
\begin{figure}[t]
    \centering
    \includegraphics[width=1\linewidth]{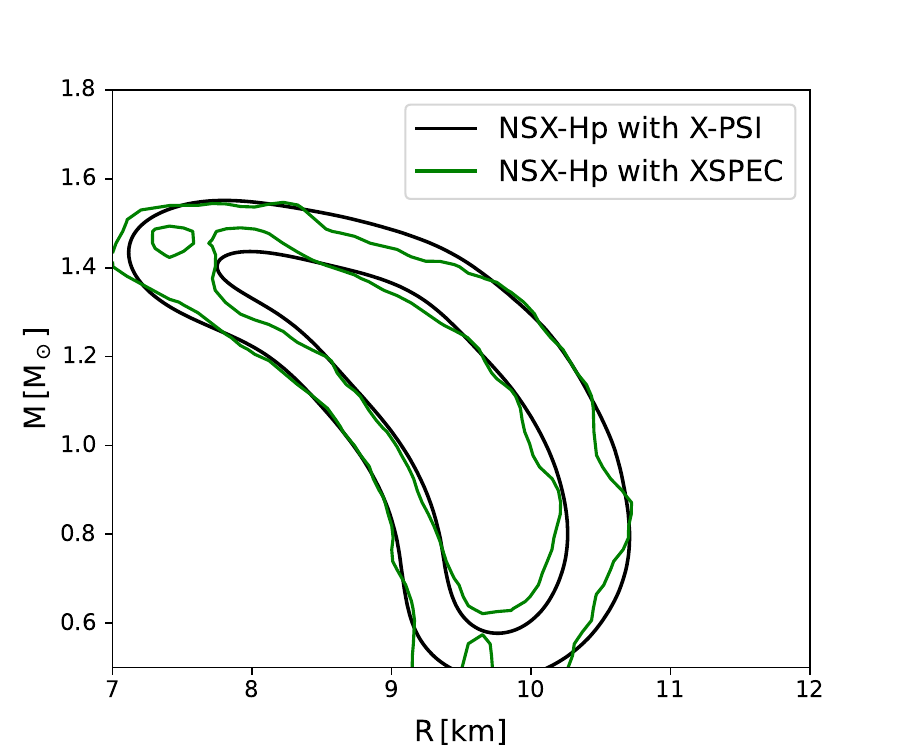}
    \caption{Posterior distribution comparison for the \texttt{nsx}-Hp model between X-PSI and XSPEC. Contours are the 68\% and 95\% credible regions (CRs) from XSPEC in green and X-PSI in black.}
    \label{fig:bogd17}
\end{figure}

Once the theoretical models were confirmed to be the same, we performed a comparative inference with the \texttt{nsx}-Hp model with both \texttt{XSPEC} and X-PSI using a small subset of the observations: the 180\,ks of ACIS-S data from 2014 -- 2015. As was mentioned above, the \texttt{XSPEC} implementation of NS atmosphere models assumes a nonrotating NS. In X-PSI, we applied a small spin frequency, $f=10\,$Hz, which closely reproduces the nonrotating case. Indeed, at 10\,Hz, for a 1.4\,\msun\ 12\,km NS, the radii at the poles and equator differ by only 27\,cm, and the maximum photon energy difference between the blueshifted and redshifted sides at the equator is $\Delta E=0.5$\%, for a star seen edge-on (from Eqs.~15 and 22 of \citealt{bogdanov_constraining_2019_paperII}).
Using \citetalias{baubock_rotational_2015}'s Eq. (27), the expected bias on \rinf\ is $<1\times 10^{-5}$, which is negligible compared to the precisions of the software and of other biases involved. 

Interstellar attenuation and pileup effects were added in both inferences using, respectively, the \texttt{Tbabs}\footnote{\url{https://pulsar.sternwarte.uni-erlangen.de/wilms/research/tbabs/}} model and the pileup module. In X-PSI, \texttt{Tbabs} is in the form of pre-calculated tables of energy-dependent attenuation (e.g., from \citealt{salmi_atmospheric_2023}), while the pileup is the module described above. 

We obtained the contours shown in Fig.~\ref{fig:bogd17}, which compare the \mr\ posteriors obtained with X-PSI (via nested sampling) and with XSPEC (using a Markov chain Monte Carlo sampling). The excellent match confirms that the setup developed in X-PSI for qLMXBs is correct, with this advantage of including a number of effects, such as rotation, that were neglected in XSPEC.

The model \texttt{nsx}-Hp will not be used further for the present analyses, since the temperatures of NS atmospheres in qLMXBs reach $10^6$\,K, i.e., they are thus fully ionized at the most important depths. All analyses in the remainder of this paper were performed with the \texttt{nsx} fully ionized hydrogen model.

\section{Application to 47~Tuc X7 -- Model setup and results}
\label{sec:results}
\subsection{Models and priors}
\label{sec:results_priors}
We used X-PSI to analyze the entire set of \textit{Chandra} data described in Sect.~\ref{sec:data_Chandra}. Several models were used to investigate the effects of the different biases described in Sect.~\ref{sec:biases}.

The \texttt{Everywhere} model corresponds to a uniform emission at a temperature, $T_{\rm{surf}}$, from the entire surface, i.e., the equivalent to what has been done with \texttt{XSPEC} in the literature when setting ``norm=1'' (fraction of surface emitting being the whole surface) with \texttt{nsatmos} or \texttt{nsx}. 
Adding hot spots to the model is done with the \texttt{HotRegion} module in X-PSI. A uniform circular hot spot of temperature $T_\mathrm{spot}$ can be described by a single temperature model, \texttt{ST}, which can be coupled with an \texttt{Elsewhere} instance to represent the emission from the rest of the surface at temperature $T_\mathrm{else}$. As is demonstrated below, the data do not require the addition of a second hot spot.

The atmosphere model used is the fully ionized hydrogen \texttt{nsx}-H atmosphere, as the companion is believed to be hydrogen-rich (see Sect.~\ref{sec:47Tuc_desc} for details). The \texttt{Tbabs} interstellar absorption model is also included. Both these models are the same as what has been used in most X-PSI analyses of MSPs.

Except when stated otherwise, priors on all parameters are uniform between their respective bounds. The mass was set to vary between 1.0 and 3.0 \msun. Note that \citet{bogdanov_neutron_2016} used a $0.5\,\msun$ lower limit. Our choice stems from the oblateness approximation used in X-PSI only being accurate down to masses of 1\,\msun\ \citep{algendy_universality_2014} for our allowed range of radii. The equatorial radius varies between 3$r_g(M)$ ($\sim$4.5\,km for a 1.0\,M$_\odot$ NS), where $r_g(M)$ is the Schwarzschild radius, and 16.0\,km. Additional priors were added on both \mns\ and \rns\ by the compactness, preventing the polar radius from being smaller than the photon sphere, i.e., $R_\mathrm{polar} > 3r_g(M)$, and by keeping the surface gravity $g \approx (1+z) G\mns/\rns^2$ within the bounds of the pre-calculated atmosphere model table: $13.7 \leq \log(g) \leq 15.0$. 

For the distance to 47~Tuc, we adopted the measurement of $d=4.52\pm0.03$\,kpc \citep{baumgardt_accurate_2021} as a Gaussian prior in the inference. Except when mentioned otherwise, the logarithm of the temperature varied between 5.1 and 6.8 (corresponding to the bounds of the \texttt{nsx}-H atmosphere model table). In previous works, the column density, $N_H$, was estimated to be around $\left(4.0-5.0\right)\times10^{20}\,\rm{cm}^{-2}$ in \citet{heinke_hydrogen_2006} but $<3\times10^{20}\,\rm{cm}^{-2}$ in \citet{bogdanov_neutron_2016}. We defined a conservative uniform prior between 0 and 1$\times10^{21}\,\rm{cm}^{-2}$.

Due to molecular contamination on the surface of the ACIS detector\footnote{\url{https://cxc.cfa.harvard.edu/cal/memos/contam\_memo.pdf}}, the effective area at low energies decreased significantly (circa 2010s). The low energy gain calibration and energy distribution are also uncertain. Because of these calibration issues, we restricted the energy range to 0.5 -- 4.0\,keV, where data are less affected, for all spectra acquired in 2014 and after. To account for possible uncertainties of the absolute flux calibration in this energy range, we defined an energy-independent scaling factor, $C_{\rm{HE}}$, with a Gaussian prior of 1.00$\pm$0.03. 
For data taken in the 2000s, the energy range 0.3 -- 4.0\,keV was used, since molecular contamination was less important at the beginning of the mission. However, because of known calibration uncertainties in the range 0.3 -- 0.5\,keV, we employed a second, broader, scaling factor, $C_{\rm{LE}}$, with a Gaussian prior of 1.0$\pm$0.1. 

Since pileup is a known and non-negligible source of bias even for small fractions of pileup (see Sect.~\ref{sec:pileup_bias}), it was implemented in all the models using the module presented in Sect.~\ref{sec:pileup}. Only the grade migration parameter, $\alpha$\footnote{Note that in this work the parameter $\alpha$ corresponds to the pileup, contrary to most other X-PSI-related works where $\alpha$ corresponds to the energy-independent scaling factor, here named $C$.}, was left free, with a uniform prior between 0 and 1. The frame time was set for each dataset according to Table~\ref{tab:data}, while the other parameters were set to the same default values as in \texttt{XSPEC}. 

The effects of rotation discussed in Sect.~\ref{sec:spin} depend on the NS spin frequency. While it is precisely known for MSPs, this is not the case of qLMXBs. As a consequence, $f$ is a free parameter in some of our models, with a uniform prior between 1 and 700\,Hz. This corresponds to spin periods from 1.4\,ms to 1\,s, a conservative range for NSs in the process of recycling through accretion. The inclination angle, $i$ (or equivalently, $\cos i$), between the rotation axis and the line of sight is also unknown and given a uniform prior on $\cos i$ between 0 and 1.

When adding a hot spot, several geometrical parameters are needed to describe the spot colatitude, $\theta_{\rm{spot}}$, phase, $\phi_p$, and spot angular radius, $\zeta_{\rm{spot}}$. The colatitude, $\theta \in [0,\pi]$, has a prior such that $\cos\theta$ is uniform, and the phase, $\phi$, is between $-$0.25 and 0.75, which allows it to cover the entire surface. The spot radius is uniform between 0 and $\pi/2$, such that a hot spot can cover at maximum half of the NS.

\subsection{Accounting for sources of biases}
All runs described below were performed using X-PSI v3.1.1, with the Multinest/PyMultinest sampler \citep{feroz_multinest_2009, buchner_pymultinest_2014}. We used $10^4$ live points, with a sampling efficiency of 0.01 (a low value needed to properly sample curved and elongated posteriors, as shown by \citealt{dittmann_notes_2024}, such as the \mr\ posteriors of qLMXBs), and an evidence tolerance of 0.1. Most of these parameters are identical or similar to values used in previous X-PSI analyses (e.g., \citealt{mauviard_nicer_2025, vinciguerra_updated_2024}).

\subsubsection{\texttt{Everywhere} with a fixed frequency -- Default setup}
\label{sec:EWfixF}
\begin{figure}
    \centering
    \includegraphics[width=1\linewidth]{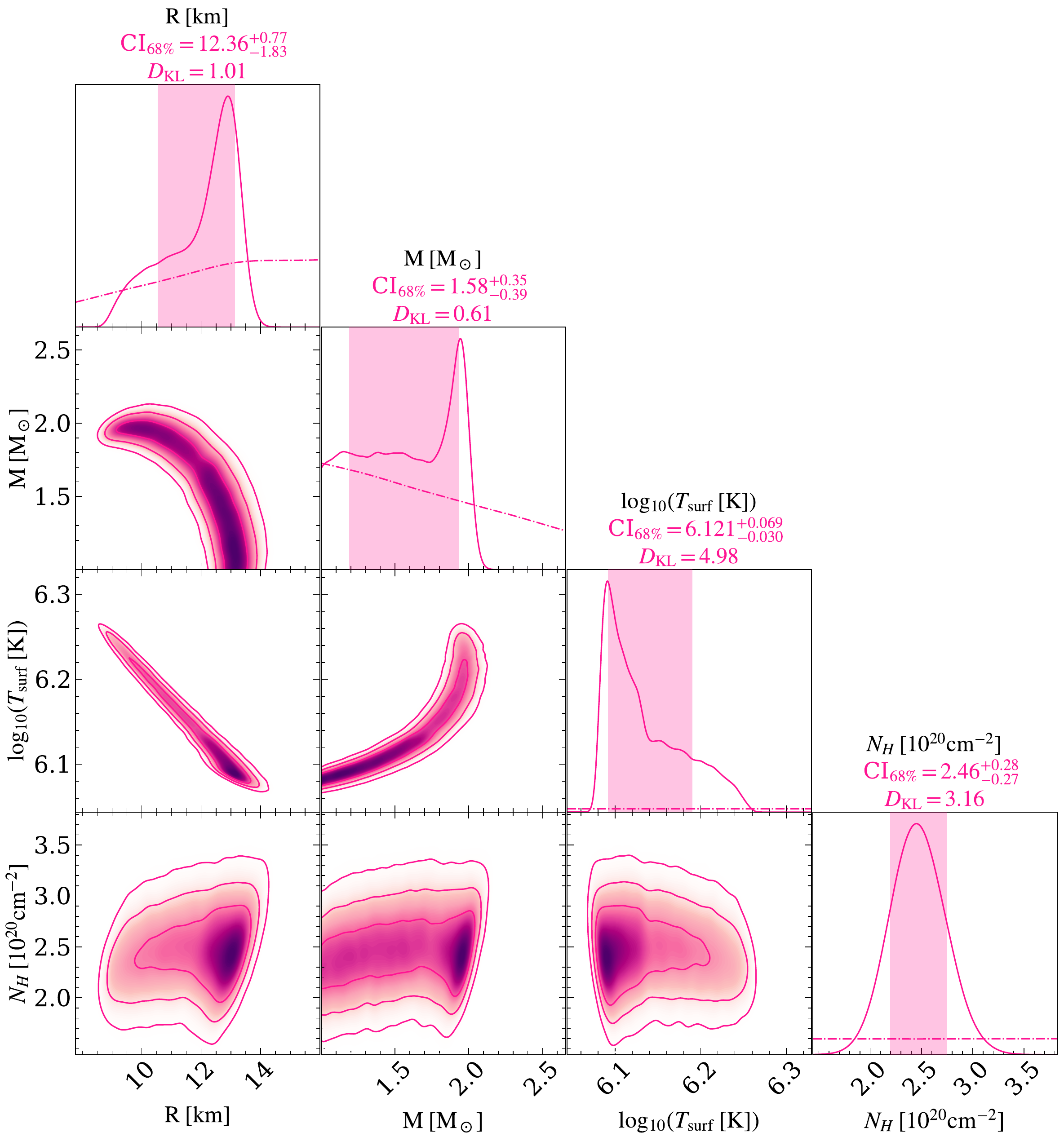}  
    \caption{Posterior distribution of an \texttt{Everywhere} model with fixed $f=10$\,Hz. The marginal distributions of \mns, \rns, \texttt{Everywhere} temperature $T_{\rm{surf}}$ and $N_H$ are along the diagonal (solid lines), while the joint posteriors with 68.3, 95.4, and 99.7\% CRs are off-diagonal. The priors are represented in dash-dotted lines on the marginal distributions panels. The medians and 68\% credible intervals (CIs) are indicated at the top of the panels and are represented by the shaded vertical bands in the marginal distribution panels.}
    \label{fig:EWfixF_MR}
\end{figure}

We first explored the \texttt{Everywhere} model, with a fixed frequency of $f=10$\,Hz, to reproduce as closely as possible qLMXB analyses in the literature. The low spin frequency chosen minimized the effects of rotation and oblateness. This is our “default” model and allows for easy comparisons to previously published results. The posterior distributions for \mns, \rns, $T_{\rm surf}$, and $N_H$ are shown in Fig.~\ref{fig:EWfixF_MR}. The posterior distributions of all parameters can be found in Fig.~\ref{fig:full_posteriors}. 

The surface temperature is, as expected, highly correlated to \mns\ and \rns, because of the dependence of $T_\infty$ on the gravitational redshift, $z$ (see Sect.~\ref{sec:intro}). Its median is $\log_{10} (T / [K]) = 6.12$, with a main peak at 6.1 and a tail extending up to $\sim 6.25$. It is consistent with \citet{bogdanov_neutron_2016} and other works that reported effective temperatures of around 100 -- 110\,eV, corresponding to $\log_{10}\,(T / [\rm{K}])=6.06-6.1$. The column density, $N_H$, is well constrained from our fit, with a value of $\left(2.5 \pm 0.3\right)\times 10^{20}\,\rm{cm}^{-2}$, in agreement with the upper limit of \citet{bogdanov_neutron_2016}.

To investigate the relevance of using the low energy range 0.3 -- 0.5\,keV and its associated scaling factor, an inference was performed with the entire dataset restricted to the energy range 0.5 -- 4.0\,keV. The notable difference comes from a column density that is less constrained ($N_H = 2.8 \pm 0.7 \times 10^{20}\,\rm{cm}^{-2}$ with this restricted energy range), leading to broader constraints on \mns\ and \rns\ by 37\%. Despite its poorer calibration, we deemed it useful to add the low energy range, which will be used for the rest of this work.

\subsubsection{\texttt{Everywhere} with a free frequency -- Adding rotation}
\label{sec:EWfreeF}
\begin{figure}
    \centering
    \includegraphics[width=1\linewidth]{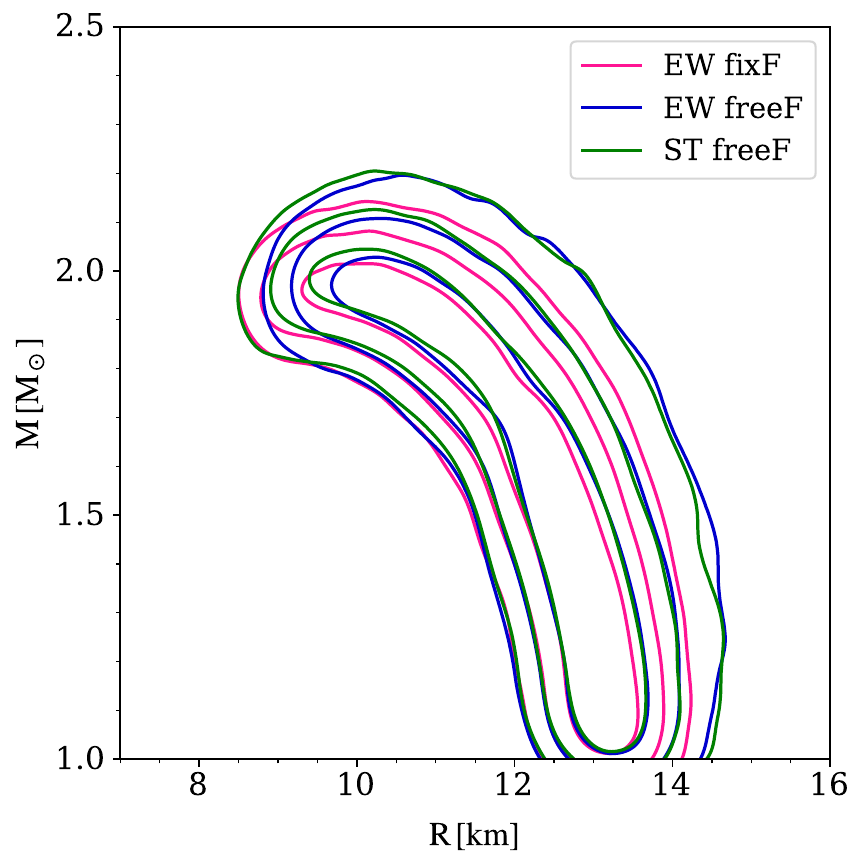}
    \caption{2D \mr\ posterior distributions with the three scenarios explored in this work: the default \texttt{Everywhere} model with fixed $f$ in pink and free $f$ in blue, and the \texttt{ST} model with free $f$ in green.}
    \label{fig:compa_MR_3models}
\end{figure}

\begin{figure}
    \centering
    \includegraphics[width=1\linewidth]{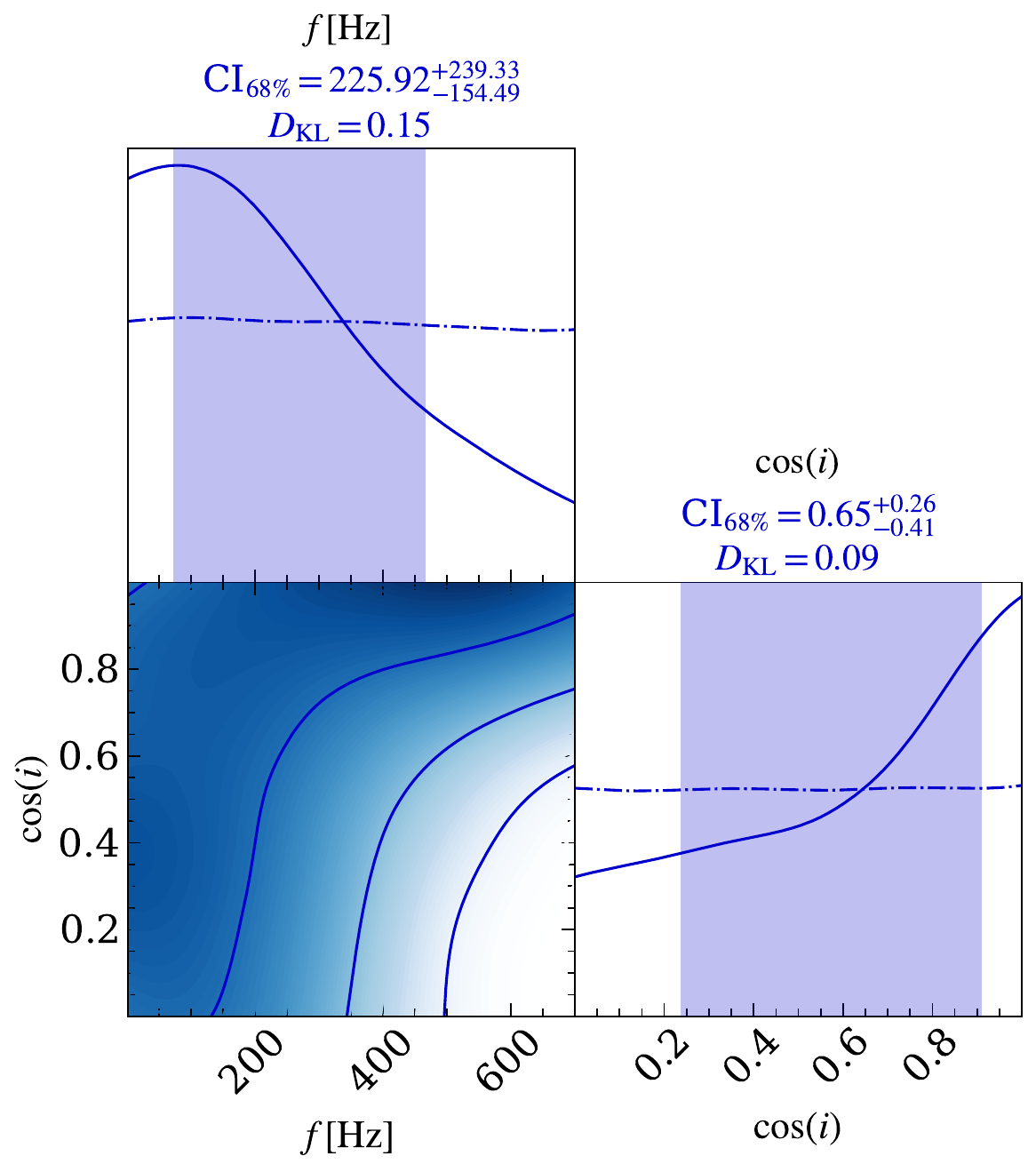}
    \caption{Posterior distributions for $f$ and $\cos i$ of the \texttt{Everywhere} model with free $f$. As in Fig.~\ref{fig:EWfixF_MR}, the joint posterior is in the bottom left while the marginal distributions and associated priors (dash-dotted lines) and 68\% CI are in the diagonal panels.}
    \label{fig:EWfreeF_Fcosi}
\end{figure}

Once the default model has been set, the first source of uncertainty to include is the effect of rotation. By freeing the spin frequency, $f$, as well as the inclination angle, $i$, we include in the modeling the effects of the Doppler shifts and the oblateness on the spectra. The \mr\ posterior distribution with this model is shown in blue in Fig.~\ref{fig:compa_MR_3models}, on top of the distribution from the default (in pink). The $f$ -- $\cos i$ posterior distribution is shown in Fig.~\ref{fig:EWfreeF_Fcosi}. The posterior distributions of all parameters can be found in Fig.~\ref{fig:full_posteriors}.

The \mr\ 2D distributions from this model (“EW freeF”) and the previous one (“EW fixF”) yield consistent results. As expected, the \mr\ constraints are broader when $f$ and $i$ are free, with the 68\% CIs at 1.4\,\msun\ increasing by 14\%, since we include a source of uncertainty in the form of two extra (unconstrained) parameters. 
This is in agreement with simulations that predicted an increase of up to around 10\% \citetalias{baubock_rotational_2015}. The temperature and column density posteriors stay very similar to the default.

Both $f$ and $i$ were unconstrained by the spectra (see Fig.~\ref{fig:EWfreeF_Fcosi}), which is not surprising since the Doppler shift is subtle ($\Delta E \leq 0.5$\%). 
Nonetheless, the sampling seems to reject the high-inclination (low $\cos i$) -- (high $f$) parameter space. This suggests that the spectrum does not require significant broadening due to the Doppler effect. The low frequencies ($f\lesssim 300$\,Hz) of all $\cos i$ values are almost fully included in the 95\% CRs, further strengthening this conclusion, as low frequencies also correspond to lower broadening.

\subsubsection{\texttt{ST}+\texttt{elsewhere} with a free frequency -- Adding rotation and hot spots}
\label{sec:STfreeF}
By using a \texttt{ST} component coupled with an \texttt{Elsewhere} instance, we included the second source of uncertainties due to possible surface inhomogeneities.
In this model (our \texttt{ST} model, hereafter), the prior on both temperatures was adjusted compared to runs with \texttt{Everywhere} by restricting to $\log_{10} (T / [\rm{K}])>6.0$. This stems from the results of the \texttt{Everywhere} model showing that the surface temperature is $\log_{10}(T / [\rm{K}])\sim 6.1$. In addition, \citet{elshamouty_impact_2016} found a pulsed fraction 90\% upper limit of 16\%, which suggests that if a hot spot is present, its temperature cannot be significantly different from the rest of the surface ($\Delta T \leq 0.055$\,keV).
Sampling temperatures below $10^6$\,K explored additional modes where the hot spot would be very large ($\zeta_\mathrm{spot} \geq 45\degr$) and “hot” ($\log_{10} (T_\mathrm{spot} / [\rm{K}])\sim 6.1 - 6.2$), accounting for the bulk of the emission, and where the rest of the surface would then be colder ($\log_{10} (T_\mathrm{else} / [\rm{K}])\sim 5.5$) and barely contribute to the spectrum. The reverse scenario, where the rest of the surface is “hot” and the spot is “cold,” would also be explored.
To suppress these modes, which we know from \citet{elshamouty_impact_2016} are excluded, two options were considered: either restricting the size of the spot, $\zeta$, to be small ($\zeta \leq 45\degr$) or restricting the temperature range to force both the hot spot and the rest of the surface to contribute significantly to the spectrum. We chose the latter to avoid unnecessarily being biased toward small spots.

The \mr\ posterior distribution of this scenario is shown in green in Fig.~\ref{fig:compa_MR_3models}, on top of the distribution from the other two. The posterior distributions of all parameters can be found in Fig.~\ref{fig:full_posteriors}, and the medians and 68\% CIs of the various parameters are reported in Table~\ref{tab:allparams_values}. The posteriors of Fig.~\ref{fig:compa_MR_3models} do not look significantly different from the contours of the \texttt{Everywhere} model with free $f$. Frequency and hot spot parameters are unconstrained but do not broaden the \mr\ posteriors compared to the free $f$ alone. Similarly to the uniform temperature model, little Doppler shift is required by the spectrum, as the region of (low $\cos i$) -- (high $f$), corresponding to high Doppler shifts, is excluded by the sampling. The column density is still well constrained at a value compatible with that of the previous scenarios. 

\begin{table}
    \centering
    \begin{threeparttable}
    \caption{Posterior distribution parameters of the \texttt{ST+Elsewhere} model with a free frequency. }

    \begin{tabular}{c|c}
        Parameter Name & Value from the posteriors \\
        \hline 
        \multicolumn{2}{c}{Physical parameters} \\
        mass $[\mathrm{M}_\odot]$ & $\mathrm{M} = 1.57^{+0.38}_{-0.37}$ \\
        radius [km] & $\mathrm{R} = 12.49^{+0.74}_{-1.84}$ \\ 
        radius at infinity [km] & $\mathrm{R_\infty} = 15.64^{+0.49}_{-0.45}$ \\ 
        distance [kpc] & $d = 4.521^{+0.028}_{-0.028}$ \\ 
        surface temperature [K] & $\mathrm{log}_{10}(T_{\mathrm{surf}}) = 6.120^{+0.071}_{-0.028}$ \\ 
        frequency  [Hz] & f = $244^{+248}_{-164} $ \\
        cos inclination & $\mathrm{cos}(i) = 0.73^{+0.20}_{-0.43}$ \\
        spot phase shift [cycles] & $\phi_{p} = 0.25^{+0.38}_{-0.37}$ \\
        spot colatitude [rad] &$ \Theta_{\mathrm{spot}}  = 1.85^{+0.65}_{-0.80}$ \\
        spot radius [rad] & $\zeta_{\mathrm{spot}} = 0.37^{+0.43}_{-0.27}$ \\
        spot temperature [K] & $\mathrm{log}_{10}(T_{\mathrm{spot}}) = 6.124^{+0.103}_{-0.072}$ \\
        column density $[10^{20} \mathrm{cm}^{-2}]$ & $N_H = 2.46^{+0.27}_{-0.26}$ \\ [0.15cm]
        \multicolumn{2}{c}{ACIS-S Parameters} \\
        high energy scaling factor & $C_{\mathrm{HE,\:ACIS-S}} = 1.000^{+0.028}_{-0.028}$ \\  
        low energy scaling factor & $C_{\mathrm{LE,\:ACIS-S}} = 1.000^{+0.092}_{-0.093}$ \\  
        grade migration $\tau = $ 0.4s & $\alpha_{\tau = 0.4s} = 0.22^{+0.15}_{-0.13}$ \\ 
        grade migration $\tau = $ 0.8s & $\alpha_{\tau = 0.8s} = 0.21^{+0.11}_{-0.10}$ \\ 
        grade migration $\tau = $ 3.0s & $\alpha_{\tau = 3.0s} = 0.07^{+0.10}_{-0.05}$ \\
        grade migration $\tau = $ 3.1s & $\alpha_{\tau = 3.1s} = 0.55^{+0.02}_{-0.01}$ \\[0.15cm]
        \multicolumn{2}{c}{ACIS-I Parameters} \\
        high energy scaling factor  & $C_{\mathrm{HE,\:ACIS-I}} = 1.000^{+0.028}_{-0.028}$ \\ 
        low energy scaling factor & $C_{\mathrm{LE,\:ACIS-I}} = 0.999^{+0.092}_{-0.092}$ \\  
        grade migration $\tau = $ 0.9s & $\alpha_{\tau = 0.9s} = 0.11^{+0.14}_{-0.08}$ \\
        grade migration $\tau = $ 0.5s & $\alpha_{\tau = 0.5s} = 0.34^{+0.34}_{-0.24}$ \\ 
        grade migration $\tau = $ 3.2s & $\alpha_{\tau = 3.2s} = 0.53^{+0.04}_{-0.04}$ \\
        \hline
    \end{tabular}
    \label{tab:allparams_values}
    \tablefoot{Medians and 68\% CIs for the posterior distribution of the \texttt{ST+Elsewhere} model with a free frequency. Parameters and symbols are as mentioned in the text. Grade migration parameters for each of the grouped observations from Table~\ref{tab:data} are indicated with the associated frame time value.}
    \end{threeparttable}
\end{table}

\begin{figure}[t]
    \centering
    \includegraphics[width=1\linewidth]{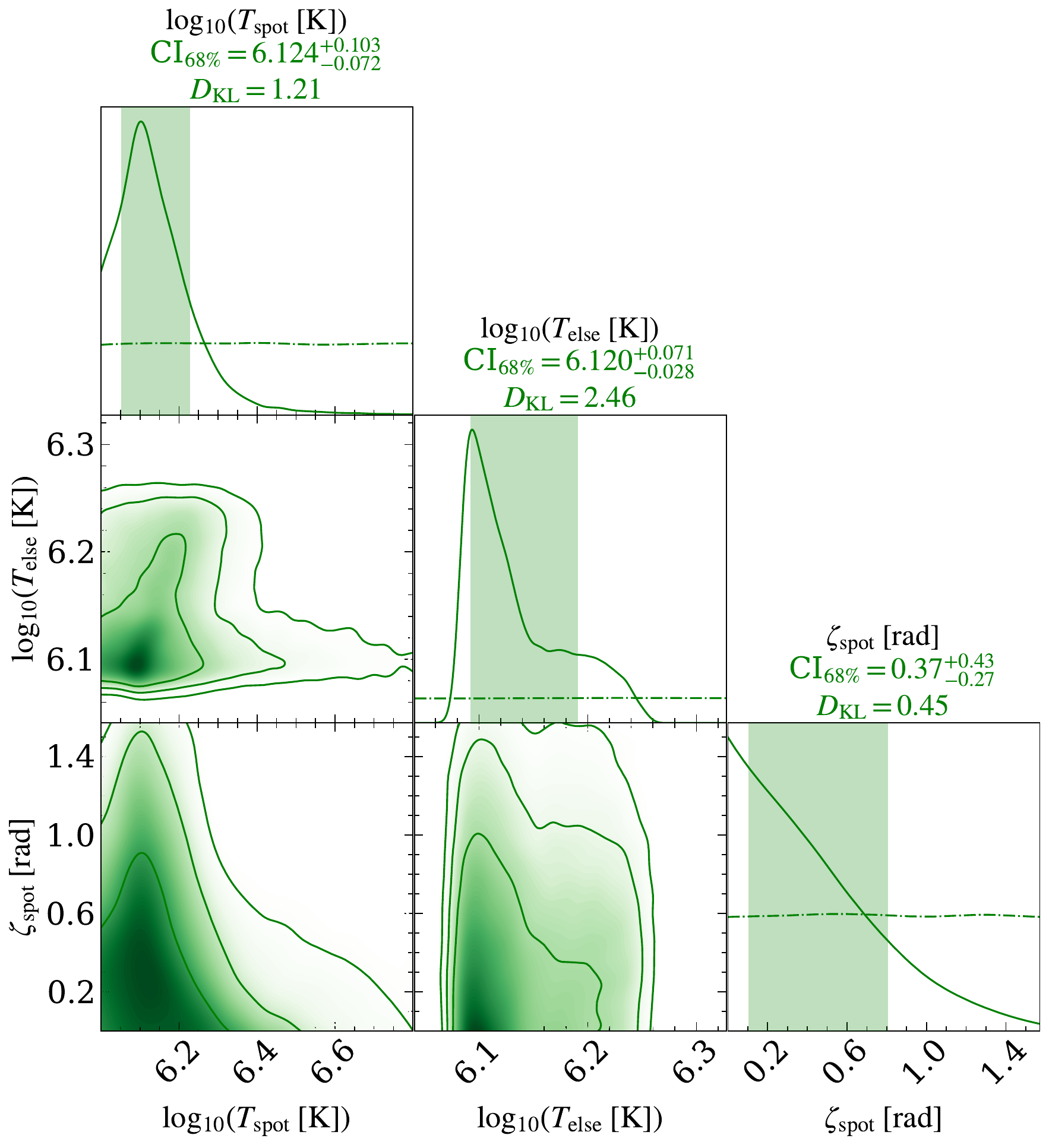}
    \caption{Posterior distributions for the angular size of the hot spot, $\zeta_\mathrm{spot}$, and temperatures of the hot spot, $T_\mathrm{spot}$, and the rest of the surface, $T_\mathrm{else}$, of the \texttt{ST} model. The joint posteriors are in the bottom left panels, while the marginal distributions are on the diagonal, with their associated priors (dash-dotted lines) and 68\% CI.}
    \label{fig:STfreeF_params}
\end{figure}

The distributions for $T_\mathrm{spot}$ and $T_\mathrm{else}$ were left free to sample the whole range of $\log_{10}(T / [\rm{K}])$ from 6.0 to 6.8, and their posterior distributions are shown in Fig.~\ref{fig:STfreeF_params}. $T_\mathrm{else}$ is well constrained, with little to no deviation from the \texttt{Everywhere} temperature distributions of previous scenarios. $T_\mathrm{spot}$ explored the entire allowed range, and its resulting median is slightly higher than that of $T_\mathrm{else}$. However, the 2D posterior distribution shows that sampling concentrated along the region where $\log_{10} (T_\mathrm{spot}/ [\rm{K}]) = \log_{10} (T_\mathrm{else} / [\rm{K}])$, noticeable in the 68\% CR. In other words, in this scenario, there is a preference for the two temperatures to be equal or very close to one another. This corresponds to the \texttt{Everywhere} scenarios, which likely explains why the \mr\ posteriors did not change. Regions of the parameter space where temperatures are different were nonetheless explored, mostly along a horizontal line at $T_\mathrm{else}\sim10^{6.1}$\,K (see Fig.~\ref{fig:STfreeF_params}), corresponding to an almost unconstrained $T_\mathrm{spot}$. 
We note that, as $T_\mathrm{spot}$ increases beyond $10^{6.3}$~K, the spot size decreases to minimize its contribution to the spectrum.

We also tested a different hot spot model, the concentric single temperature (\texttt{CST}) model, which describes an annulus of temperature $T_\mathrm{spot}$ and radius $\zeta_1$, with a smaller concentric radius of radius $\zeta_2 < \zeta_1$ that is omitted, i.e., that will not have temperature $T_\mathrm{spot}$, but instead will be filled by the \texttt{Elsewhere} instance (or simply not emit -- hence omitted -- if there is no \texttt{Elsewhere}). This model aimed at investigating hot spot geometries that would allow one to sample annuli, which, depending on the geometry, do not necessarily produce pulsations. This new geometry configuration yielded identical results to the \texttt{ST} results. No additional constraints were provided.

\section{Discussion}
\label{sec:discussion}

\subsection{Comparison of our scenarios}

\begin{table}[t]
    \centering
    \begin{threeparttable}
    \caption{Log-evidence and maximum likelihood values for the different models.}
    \begin{tabular}{c|c|c|c}
        Scenarios & $\ln \mathcal{Z}$ & $\ln p( \text{d} | \theta_{\text ML})$ & $\chi^2$/dof\\
        \hline 
        \texttt{EW} fixed F  & $-21832.47$ & $-21806.04$ & 258.6/238\\
        \texttt{EW} free F & $-21832.45$ & $-21806.14$ & 263.5/236\\
        \texttt{ST} free F & $-21830.88$ & $-21806.06$ & 261.3/232\\
        \hline
    \end{tabular}
     \label{tab:evidence}
     \tablefoot{$\chi^2$ values and degrees of freedom are only provided for reference, but are to be taken with care as they do not come from a minimization and since the data are un-binned.}
    \end{threeparttable}
\end{table}
The evidence and maximum likelihood values for each of the models presented in this work are reported in Table~\ref{tab:evidence}. 
Despite the increasing complexity of the models (adding 6 new parameters in total), the evidence stays almost identical, with $\left(\Delta \ln \mathcal{Z}\right)_\mathrm{max} = 1.6$. This is not significant enough to confidently favor one model over another, according to \citet{kass_bayes_1995, vinciguerra_x-psi_2023}. Since there are no detected pulsations in this source (pulsed fraction $<8.5$\%; see Sect.~\ref{sec:PF}) and no obvious second thermal component in the spectrum, it is not surprising that the \texttt{ST} model does not perform better than an \texttt{Everywhere} model. The purpose of this work was not to find the exact geometric configuration of the emitting region, but to incorporate sources of uncertainties in the inference to ensure that the \mr\ contours are not biased. 

Using more complex geometries of hot regions, or incorporating two (or more) hot spots instead of a single one, could have been explored. However, adding a single circular hot spot does not improve the fit and could be argued to be overfitting, despite it being done with the aim of providing a bias-free model. Adding a single hot spot also makes the model more complex and have much longer inference runs (\texttt{ST} was 7 times longer than both \texttt{Everywhere} runs). More emitting regions would only increase those two aspects but would not provide more information than the current results. 

A comparison with the results from \citet{bogdanov_neutron_2016} was also performed. In their analysis, they found a radius in the 10.4 -- 11.9\,km range (68\% CI) for a 1.4\,\msun\ NS, while, for this same mass, we find 12.5 -- 13.3\,km with our most conservative \texttt{ST} model\footnote{This value is calculated using a horizontal slice at 1.4~\msun\ in Fig.~\ref{fig:compa_MR_3models}}. A direct comparison of Fig.~\ref{fig:compa_MR_3models} with their \mr\ contours (Fig.~7 in \citealt{bogdanov_neutron_2016}) is, however, not relevant because of the different datasets, different atmosphere models, different processing and calibration, and more importantly, a different implementation (ray tracing, oblateness, frequency, hot spots, etc.). Trying to reproduce their configuration as closely as possible, with the same datasets and our default model, yielded comparable contours (68\% CIs of 9.9 -- 11.6\,km for a 1.4~\msun\ NS), with the slight shift to lower radii attributed to the unavoidable differences mentioned above.

\subsection{The effects of rotation}
\label{sec:rotation_Baubock}
\citetalias{baubock_rotational_2015} studied the effects of rotation on a blackbody spectrum using the Hartle-Thorne metric, which incorporates the oblateness of the star as well as the quadrupole moment. They estimated the effect of rotation on the flux and temperature and showed that the distortion of the spectrum can bias \rinf\ at the $\simeq$5\% level at most for a 12\,km NS spinning at f=700\,Hz. This value is, however, a general estimate based on simulations. The real value depends on a number of parameters, including \mns, \rns, and $f$ of the NS, as well as $i$. 

Several differences can be noted between the present work and \citetalias{baubock_rotational_2015}. Firstly, \citetalias{baubock_rotational_2015} considers a blackbody while we consider a NS-atmosphere model, which is spectrally broader than a blackbody spectrum \citep{zavlin_model_1996}. Secondly, the metrics used in both works are different. For this last point, \citetalias{baubock_rotational_2015} does mention, however, that the oblateness seems to be the dominant effect in their work, but the quadrupole moment or other effects (such as frame dragging or gravitational lensing) could come into play in some specific systems and geometries. In X-PSI, the use of oblateness in a Schwarzschild metric was estimated to be accurate at the 0.1\% level, compared to numerical relativity \citep{bogdanov_constraining_2019_paperII}. This is also deemed sufficient for this work. Finally, while \citetalias{baubock_rotational_2015} relied on simulations, we computed our shifts from the fitting of actual data.

\begin{figure}[t]
    \centering
    \includegraphics[width=0.9\linewidth]{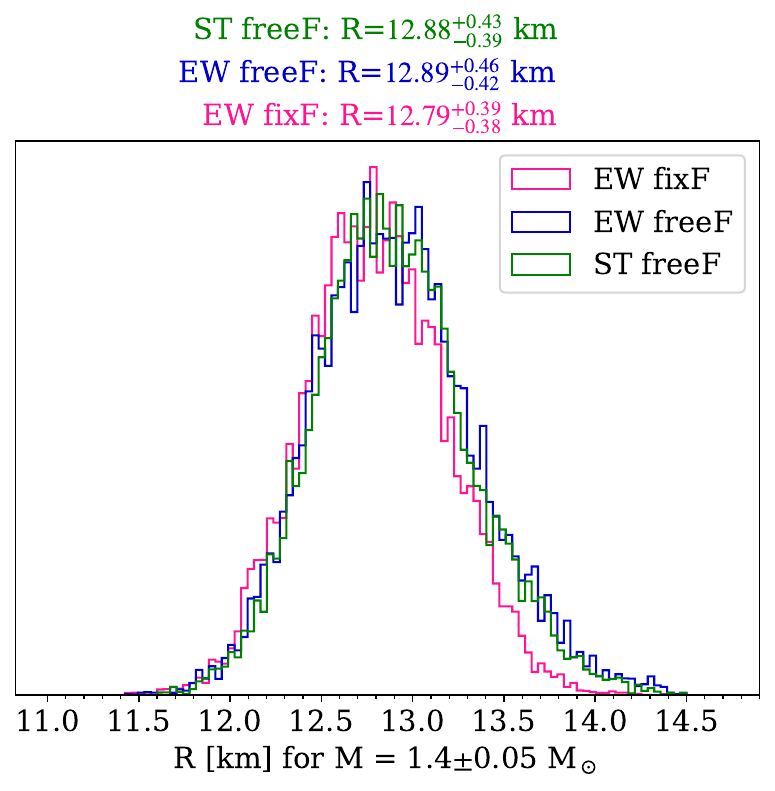}
    \caption{Histogram of the samples of our three models, selecting samples where the mass is restricted to 1.40$\pm$0.05\,\msun. Each point is weighted by its sample probability. The median and 68\% CIs are indicated at the top.}
    \label{fig:cutMR}
\end{figure}
We show in Fig.~\ref{fig:cutMR} the distribution of R posterior samples for masses around 1.4$\pm$0.05\,\msun, with the resulting medians and CIs. A shift of 0.7 -- 0.8\% is measured when comparing the medians of the default distribution and the ones where frequency is freed. The CIs also increase by about 13\% and 6\% for the \texttt{Everywhere} and \texttt{ST} models, respectively. 
Equation (27) of \citetalias{baubock_rotational_2015} predicts a bias on \rinf\ of 0.65\% for our median values for \mns=1.4\,\msun: \rns=12.9\,km, f=240\,Hz. Given that the relation has a precision of 0.5\% for spins below 800\,Hz, and given the other differences between the methods mentioned above, we are in agreement within uncertainties. 

\subsection{The pulsed fraction}
\label{sec:PF}
\begin{figure}[t]
    \centering
    \includegraphics[width=1\linewidth]{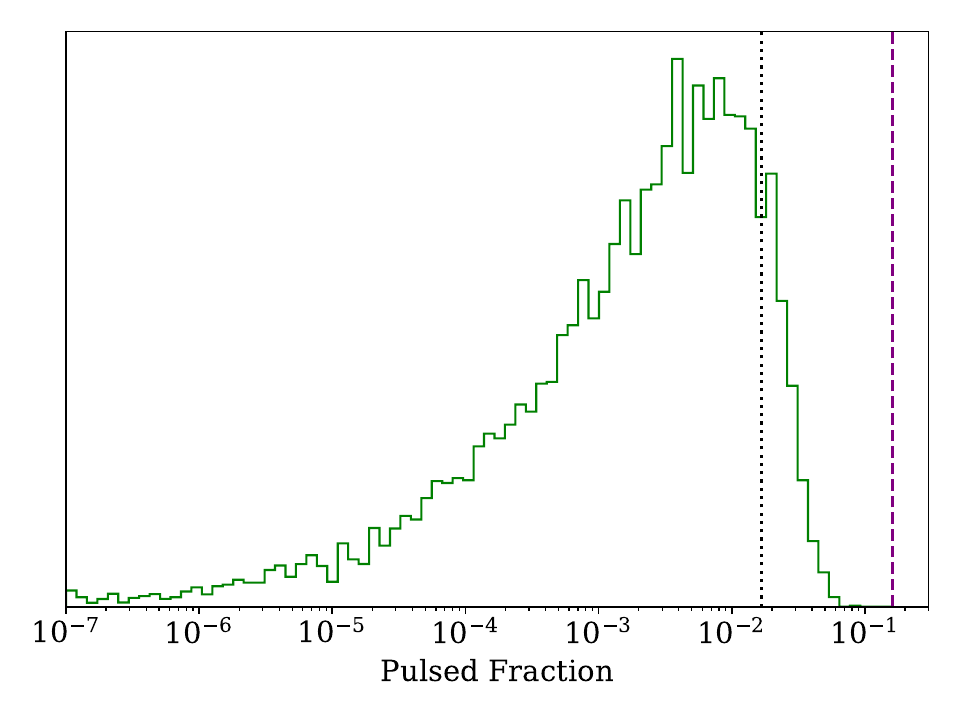}
    \caption{Histogram of pulsed fractions computed from the modeled signal of the last 10000 samples from the \texttt{ST} inference run. Only points with pulsed fractions above $10^{-7}$ are shown. The dashed purple line represents the 90\% upper limit of 16\% computed by \citet{elshamouty_impact_2016}, while the dotted black line corresponds to the one from this work.}
    \label{fig:pulse_frac}
\end{figure}

One definition of the pulsed fraction is the amplitude of the oscillation over the average flux: $PF = (F_\mathrm{max} - F_\mathrm{min})/(F_\mathrm{max}+F_\mathrm{min})$. For our \texttt{ST} model, this quantity can be computed based on the modeled signal for each of the final samples, resulting in the distribution shown in Fig.~\ref{fig:pulse_frac}.

Only samples with a pulsed fraction higher than $10^{-7}$ are shown, which excludes $\sim$10\% of them. We obtain a 90\% credible level at $PF=1.7\%$ ($PF=6\%$ at the 99.97\% credible level), and the maximum value is $PF=8.4\%$. 
\citet{elshamouty_impact_2016} computed a pulsed fraction upper limit on 47~Tuc X7 using Timing mode \textit{Chandra} HRC data in addition to the ACIS dataset of 2014 -- 2015 for spectral constraints, finding a 90\% confidence upper limit of 16\% (represented as the purple line in Fig.~\ref{fig:pulse_frac}), which is almost twice our maximum value. 

However, they used fiducial values of \mns, \rns\ and surface temperature of the NS in most of their simulations. The choice of said values could partly explain the differences, as their \rns\ and $T_{\rm surf}$ are lower than our inferred values. They also showed that the spot size and $T_{\rm spot}$, as well as $f$, have the most influence on the pulsed fraction distribution (and consequently on the upper limits). This could explain the difference between our value and theirs, as our resulting pulsed fractions are integrated over the whole parameter space. Finally, they only used a subset of ACIS-S data, while we used the entire set, leading to a much higher S/N, which could have helped place more stringent constraints on the existence of a second thermal component. Therefore, the present work suggests that the spectrum can constrain the pulsed fraction ($PF<8.5$\% here) and the presence of hot spots just as well as (or better than) blind pulsation searches.

\subsection{The remaining systematics}
Not all known sources of systematic uncertainties were included in this work. The main neglected one is the magnetic field. However, as is explained in Sect.~\ref{sec:biases}, it is believed to be sufficiently low to be safely ignored. 

In addition, the X-PSI framework incorporates several assumptions to compute the emission from the surface that are worth mentioning. General relativity is assumed, and the Schwarzschild metric is used to describe the spacetime around the NS. Rotational effects are included by using the oblate Schwarzschild approximation. A more precise metric, accounting for additional effects such as quadrupole or frame dragging, might have provided different results but would have required major changes in the X-PSI code and a much more computationally expensive inference for a 0.1\% or less precision increase (see Sect.~\ref{sec:rotation_Baubock}).

\subsection{EOS inference}

\begin{figure}[t]
    \centering
    \includegraphics[width=1\linewidth]{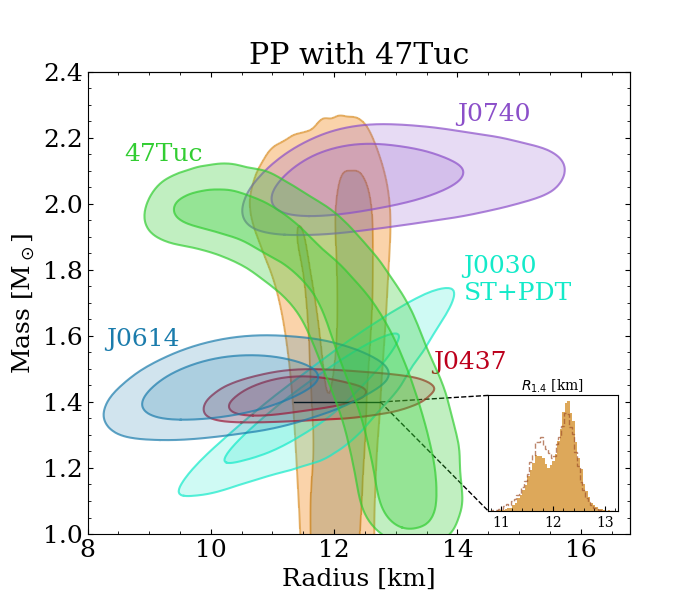}
    \caption{\mr\ posterior distributions from the EOS inference in orange. The 68\% and 95\% CRs of the \texttt{ST} model of 47~Tuc X7 are shown in green. Also shown are the CRs of the four NICER MSPs previously analyzed with X-PSI: PSR~J0740+6620 in purple, PSR~J0437$-$4715 in red, the \texttt{ST+PDT} model of PSR~J0030+0451 in cyan, and PSR~J0614$-$3329 in blue (see Sect.~\ref{sec:intro} for the references). The small inset shows the filled histogram of $R_{1.4}$ from the EOS posterior, while the dashed brown line is the corresponding histogram from \citet{mauviard_nicer_2025}.}
    \label{fig:EOS}
\end{figure}

To investigate the impact of this new robust \mr\ measurement for 47~Tuc X7 on the EOS, we used the inference software NEoST\footnote{\url{https://github.com/xpsi-group/neost}} v2.2.0 \citep{raaijmakers_neost_2025}. For conciseness, we explored only one EOS-parametrized model: a piecewise-polytropic (PP) EOS above 1.5$\rho_0$ \citep{hebeler_pp_2013} complemented by next-to-next-to-next-to-leading order chiral field theory below that density \citep{keller_cEFT_2023}. 
A more complete study with different EOS parametrized with additional \mr\ measurements from qLMXBs will be the subject of a future work.

The current approach is the same as the one presented in \citet{rutherford_constraining_2024, mauviard_nicer_2025}. The \texttt{ST} \mr\ samples of 47~Tuc X7 of the present work are added to those obtained for the four MSP \mr\ measurements\footnote{J1231$-$1411 is not used because its \mr\ samples were already informed by an EOS model \cite{salmi_nicer_2024}.} mentioned in Sect.~\ref{sec:intro}, as well as the mass-tidal deformability constraints from GW~170817 \citep{abbott_gw_2019} and GW~190425 \citep{abbott_gw_2020}. The \mr\ posteriors of the PP model are shown in Fig.~\ref{fig:EOS} along with the \mr\ measurements of 47~Tuc X7 and of the four MSPs. To quantify the improvement of the EOS constraints, we evaluated \rns\ at 1.4\,\msun, at 2.0\,\msun, the maximum possible radius, $R_{\rm TOV}$, and the maximum possible mass, $M_{\rm TOV}$, and compared them to those of \citet{mauviard_nicer_2025}, the previous analysis with the same inputs except for 47~Tuc X7. All values are reported in Table~\ref{tab:eos_params}.

\begin{table}[t]
    \centering
    \begin{threeparttable}
    \caption{Parameters deduced from the posterior EOS distributions.}
    \begin{tabular}{c|c|c}
         & \citet{mauviard_nicer_2025} & This work  \\
         \hline
         $R_{1.4}$ [km]& $12.05^{+0.56}_{-0.79}$ & $12.16^{+0.50}_{-0.78}$ \\
         $R_{2.0}$ [km]& $11.99^{+0.85}_{-1.25}$ & $12.15^{+0.76}_{-1.41}$ \\
         $R_{\rm TOV}$ [km]& $11.72^{+1.13}_{-1.33}$ & $11.97^{+0.96}_{-1.55}$ \\
         $M_{\rm TOV}$ [\msun]& $2.13^{+0.13}_{-0.18}$ & $2.13^{+0.14}_{-0.17}$ \\
         \hline
    \end{tabular}
    \label{tab:eos_params}

    \tablefoot{Comparison of parameters of this work, obtained with 47~Tuc X7, compared to those of \citet{mauviard_nicer_2025} without the 47~Tuc X7 data. $R_{1.4}$ and $R_{2.0}$ are the radii of a 1.4\,\msun\, and 2.0\,\msun\, NS, respectively. $M_{\rm TOV}$ and $R_{\rm TOV}$ are the mass and radius of a nonrotating, maximum mass NS.
    }
    \end{threeparttable}
\end{table}

Adding 47~Tuc X7 shifts the median $R_{1.4}$ to higher values by 0.9\% while shrinking the 68\% CI by 6\%. This reduction of the CIs is because, when adding PSR~J0614$-$3329, \citet{mauviard_nicer_2025} obtained a PP EOS posterior displaying a bimodality. This feature appears to be reduced when adding 47~Tuc X7, the posteriors favoring the higher radius mode. Interestingly, the $M_{\rm TOV}$ remains unchanged, while the $R_{\rm TOV}$ also shifts to higher radii without a significant reduction of the 68\% CI.

\section{Conclusions}
\label{sec:ccl}
In this work, we reintroduce qLMXBs as systems capable of producing reliable \rns\ measurements for EOS inference. To do so, we developed a new method of constraining their \mns\ and \rns\ using the software X-PSI for spectral fitting and parameter inference. This allows one to incorporate in the modeling several effects, namely rotation and surface inhomogeneity effects, that were neglected in previous works despite biasing the spectral analyses of qLMXBs, as suggested in the literature. We applied this method to the qLMXB X7 in 47~Tuc. The NS atmosphere composition may be the most important source of biases, but is known here as the companion is H-rich.

Using X-PSI, we derived \mr\ CRs for three different models, each time adding an additional source of biases: first a default uniform surface model with a fixed frequency and inclination (meant to be comparable to previous works), then freeing the frequency and inclination, and finally adding possible surface inhomogeneities. Adding these potential sources of biases allowed us to infer \mr\ posteriors that are more conservative than in previous studies, despite the additional parameters (frequency, inclination, or hot spot properties) being unconstrained. When adding the rotational effects, we measure a shift of the median radius of slightly less than a percent and an increase in the 68\% CIs by about 13\%. The posterior distributions seem to indicate that the spectrum of 47~Tuc X7 does not favor significant spectral broadening due to rotational effects. Surface inhomogeneities do not widen the CIs further, which we infer comes from the fact that the preferred region of sampling corresponds to a hot spot temperature equal to the surface temperature (i.e., an overall uniform NS surface). We obtain $\rns=12.88^{+0.43}_{-0.39}$\,km for a 1.4\,\msun\ NS (68\% CI) for our most conservative model. Using our posterior samples, we evaluated the distribution of pulsed fraction and found a 90\% upper limit of $PF=1.7\%$, i.e., a factor of 10 better than deduced from the timing mode observations with \textit{Chandra} HRC. 

We deduced from this work that the extensive dataset of 47~Tuc X7 with a more than 500\,ks exposure, providing excellent S/N, significantly helped to constrain the existence of a hot spot as a second thermal component in the spectrum, without significantly broadening the resulting \rns\ CIs. Subsequent analyses for other qLMXBs, with lower S/N than 47~Tuc X7, will demonstrate if these conclusions can be generalized. 

We also used the inferred \mr\ posteriors to evaluate the improvement of this one qLMXB measurement on the EOS, finding a 68\% CI of the PP EOS radius shrunk by 6\%. Overall, this work has demonstrated that qLMXBs can provide reliable measurements and, more importantly, can constrain the EOS on a par with the PPM method with MSPs.

\section*{Data availability}

All the data products used for this analysis as well as reproduction package and supplementary material can be found in the Zenodo repository doi:\href{https://doi.org/10.5281/zenodo.20312255}{10.5281/zenodo.20312255}. \\

\begin{acknowledgements}
    We thank the anonymous referee for their valuable input, which helped improve the clarity of this manuscript.
    We thank Anna Watts, Yves Kini, Devarshi Choudhury, Bas Dorsman, Mariska Hoogkamer, Pierre Stammler, Denis Gonzalez-Caniulef, and the X-PSI development team for the help and discussions. C.K. was partially supported through the grant EUR TESS N°ANR-18-EURE-0018 in the framework of the Programme des Investissements d'Avenir. This work was partially performed using HPC resources from CALMIP (Project 2016-19056). We acknowledge NWO for providing access to Snellius, hosted by SURF through the Computing Time on National Computer Facilities call for proposals. C.K., S.G., L.M., and N.W. are supported by the CNES and by the ANR-20-CE31-0010 (MORPHER) and ANR-25-CE31-7901-01 (DENSER) grants from the Agence Nationale pour la Recherche. T.S. acknowledges support by the Research Council of Finland grant No. 368807 and the Centre of Excellence in Neutron-Star Physics (project 374063).
\end{acknowledgements}

\bibliographystyle{bibtex/aa}
\bibliography{biblio}

\begin{appendix}

\onecolumn
\section{Full posteriors}
\begin{figure*}[ht!]
    \centering
    \includegraphics[width=1\linewidth]{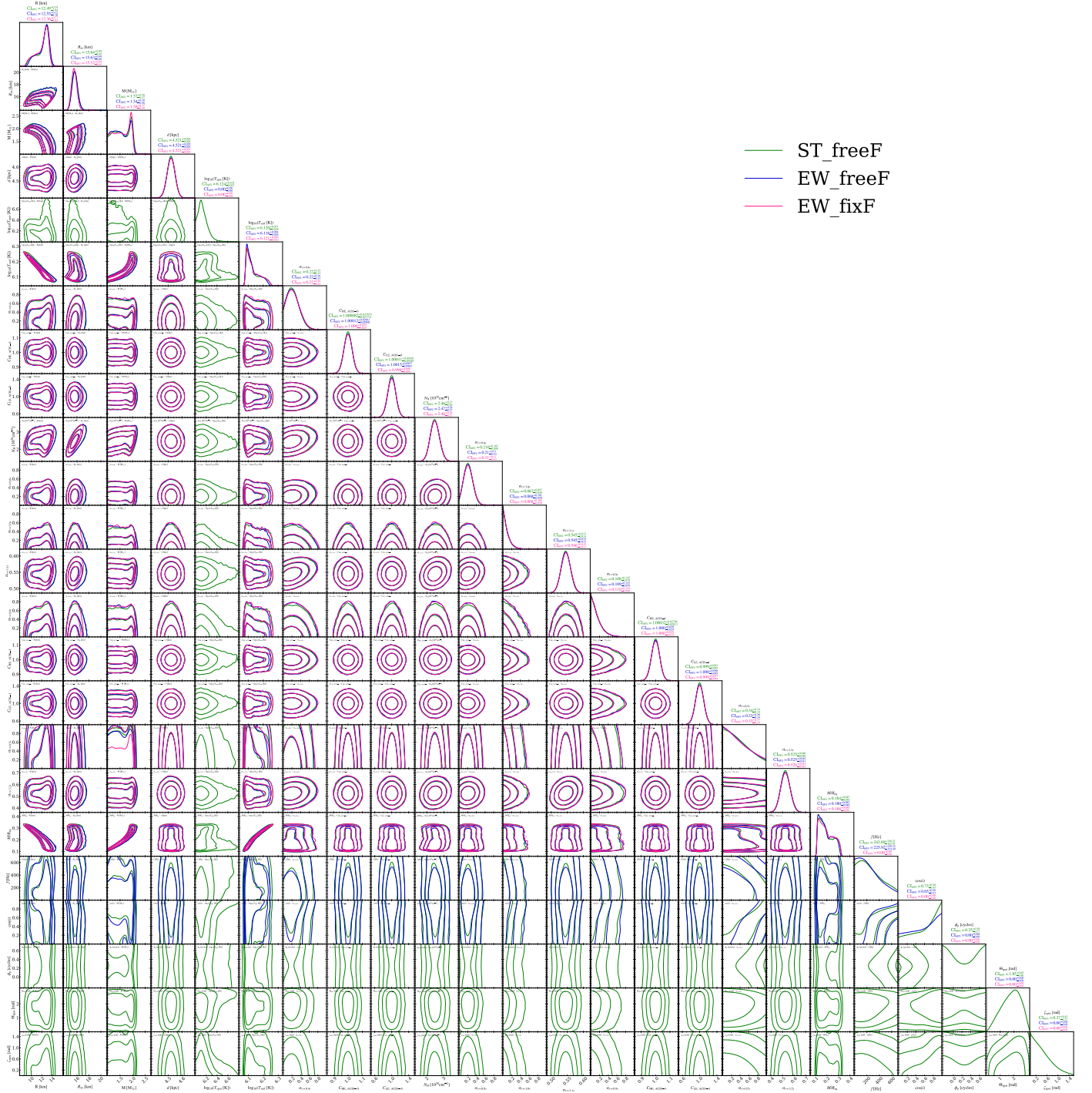}
    \caption{Posteriors of all parameters for the different models explored in Sect.~\ref{sec:results}. The joint posteriors for each pair of parameters are off-diagonal, while the marginal distributions for each parameter are along the diagonal. Each model is color-coded for visualization purposes. For comparison, the \texttt{Elsewhere} temperature $T_{\rm{else}}$ is represented on top of the \texttt{Everywhere} temperatures $T_{\rm{surf}}$. All other parameters are as mentioned in Table~\ref{tab:allparams_values} and explained in the text.}
    \label{fig:full_posteriors}
\end{figure*}

\end{appendix}

\end{document}